\documentclass[manuscript,screen]{acmart}

\AtBeginDocument{%
  }

\setcopyright{acmlicensed}
\copyrightyear{2024}
\acmYear{2024}
\acmDOI{XXXXXXX.XXXXXXX}

\usepackage{algorithmic}
\usepackage{pifont}

\usepackage{graphicx}
\usepackage{textcomp}
\usepackage{xcolor}
\usepackage[caption=false,font=footnotesize,labelformat=simple]{subfig} % caption=false is needed for IEEE template
\usepackage{lscape}

\usepackage{listings}
\usepackage{tcolorbox}
\tcbuselibrary{listingsutf8}

\usepackage{tikz}
\usetikzlibrary{shapes.geometric, arrows, positioning}

\usepackage{tabularx}
\usepackage{booktabs}
\usepackage{tabularray}
\UseTblrLibrary{booktabs}

\newcolumntype{P}[1]{>{\centering\arraybackslash}p{#1}}
\newcolumntype{M}[1]{>{\centering\arraybackslash}m{#1}}

\usepackage[utf8]{inputenc}
\usepackage{enumitem}
\usepackage{cleveref}
\usepackage{url}

\usepackage{rotating}

\usepackage[normalem]{ulem}

\definecolor{mygreen}{rgb}{0,0.6,0}
\definecolor{mygray}{rgb}{0.5,0.5,0.5}
\definecolor{mymauve}{rgb}{0.58,0,0.82}
\definecolor{backgroundgrey}{gray}{0.95}

\makeatletter
\newcommand\notsotiny{\@setfontsize\notsotiny\@vipt\@viipt}
\makeatother

\newcommand{\lstbg}[3][0pt]{{\fboxsep#1\colorbox{#2}{\strut #3}}}
\lstdefinelanguage{diff}{
  morecomment=[f][\lstbg{red!20}]-,
  morecomment=[f][\lstbg{green!20}]+,
  morecomment=[f][\textit]{@@},
}

\usepackage{dirtree}
\usepackage{mdframed}

\InputIfFileExists{preamble_tikz.tex}{}{}

\usepackage{pifont}% http://ctan.org/pkg/pifont
\newcommand{\newt}[2]{#2}

\begin{document}
%\bstctlcite{IEEEexample:BSTcontrol}

%\title{AutoChip: Can Automatic Feedback Improve the Quality of LLM-Generated Chip Designs?}
\title{Automatically Improving LLM-based Verilog Generation using EDA Tool Feedback}


\author{Jason Blocklove}
\email{jason.blocklove@nyu.edu}
\orcid{0009-0005-5619-4654}
\affiliation{%
  \institution{NYU Tandon School of Engineering}
  \city{New York}
  \state{New York}
  \country{USA}
}

\author{Shailja Thakur}
\email{st4920@nyu.edu}
\orcid{0000-0001-9590-5061}
\affiliation{%
  \institution{NYU Tandon School of Engineering}
  \city{New York}
  \state{New York}
  \country{USA}
}

\author{Benjamin Tan}
\email{benjamin.tan1@ucalgary.ca}
\orcid{0000-0002-7642-3638}
\affiliation{%
  \institution{Universary of Calgary}
  \city{Calgary}
  \state{Alberta}
  \country{Canada}
}

\author{Hammond Pearce}
\email{hammond.pearce@unsw.edu.au}
\orcid{0000-0002-3488-7004}
\affiliation{%
  \institution{University of New South Wales}
  \city{Sydney}
  \state{New South Wales}
  \country{Australia}
}

\author{Siddharth Garg}
\email{siddharth.garg@nyu.edu}
\affiliation{%
  \institution{NYU Tandon School of Engineering}
  \city{New York}
  \state{New York}
  \country{USA}
}

\author{Ramesh Karri}
\email{rkarri@nyu.edu}
\orcid{0000-0001-7989-5617}
\affiliation{%
  \institution{NYU Tandon School of Engineering}
  \city{New York}
  \state{New York}
  \country{USA}
}

\renewcommand{\shortauthors}{Blocklove et al.}

\begin{abstract}

Traditionally, digital hardware designs are written in the Verilog hardware description language (HDL) and debugged manually by engineers.
This can be time-consuming and error-prone for complex designs.
Large Language Models (LLMs) are emerging as a potential tool to help generate fully functioning HDL code, but most works have focused on generation in the single-shot capacity: i.e., run and evaluate, a process that does not leverage debugging and, as such, does not adequately reflect a realistic development process.
In this work, we evaluate the ability of LLMs to leverage feedback from electronic design automation (EDA) tools to fix mistakes in their own generated Verilog.
To accomplish this, we present an open-source, highly customizable framework, AutoChip, which combines conversational LLMs with the output from Verilog compilers and simulations to iteratively generate and repair Verilog.
To determine the success of these LLMs we leverage the VerilogEval benchmark set.
We evaluate four state-of-the-art conversational LLMs, focusing on readily accessible commercial models.
EDA tool feedback proved to be consistently more effective than zero-shot prompting only with GPT-4o, the most computationally complex model we evaluated.
In the best case, we observed a 5.8\% increase in the number of successful designs with a 34.2\% decrease in cost over the best zero-shot results.
Mixing smaller models with this larger model at the end of the feedback iterations resulted in equally as much success as with GPT-4o using feedback, but incurred 41.9\% lower cost (corresponding to an overall decrease in cost over zero-shot by 89.6\%).

% {\color{red}
% In this work, we present AutoChip, the first feedback-driven, fully automated approach to utilizing state-of-the-art LLMs to generate HDL. 
% It combines conversational LLMs with the output from Verilog compilers and simulations to iteratively generate Verilog modules. 
% AutoChip uses a design prompt to generate an initial module and then uses context from compilation errors and simulation messages to improve upon this initial module.
% We evaluate AutoChip using design prompts and testbenches from VerilogEval. 
% Results are analyzed for state-of-the-art LLMs, focusing on lightweight commercial models to demonstrate the potential improvements from feedback.
% Incorporating the most recent context from a Verilog compiler and simulator improves effectiveness over existing approaches, giving up to 87.4\% passing circuits for significantly fewer LLM queries than necessary without feedback.
% We open source our code and results.
% }
\end{abstract}





\maketitle

\section{Introduction}

Designing digital hardware with a hardware description language (HDL), such as Verilog or VHDL, is a niche skill and part of a demanding process requiring substantial expertise. 
Any mishaps can lead to implementations fraught with bugs and errors~\cite{dessouky_hardfails_2019}, and with growing demand for digital systems, there is a growing demand for techniques that can assist in generating quality HDL \newt{}{code}.
High-level synthesis (HLS) tools, for instance, are able to transform designs written in high-level software languages like C to target HDLs and implement digital hardware. 

Recent efforts have shifted the abstraction level higher, leveraging state-of-the-art Large Language Models (LLMs)~\cite{vaswani_attention_2017} to translate natural language to Verilog. 
DAVE~\cite{pearce_dave_2020} and VeriGen~\cite{thakur_benchmarking_2023} were the first efforts seeking to fine-tune LLMs specifically to generate Verilog. % in this area. % by fine-tuning models for Verilog. % over Verilog datasets.
However, VeriGen and its ilk were investigated for their use in a zero-shot manner, i.e., they \newt{}{examined} output code in response to a \newt{}{single} \newt{prompt}{prompting}. %, the combination of which is evaluated for correctness.
However, designing hardware in the real world does not work this way---code is rarely correct on the first try.
Instead, hardware designers iterate over their designs, using feedback from simulation and synthesis tools to identify and fix bugs so that an implementation will meet design specifications. 
% In other words, the HDL code will be \textit{refined} over multiple iterations. 
This feedback-based approach is \textbf{not} well reflected in existing code-generation LLMs.
Recent work~\cite{blocklove_chip-chat_2023} has proposed an iterative and interactive, conversational (or \textbf{chat}-based) approach for Verilog code generation, more closely mimicking the design flow of a human hardware engineer. 
In this case, though, feedback comes entirely from a human developer who inspects the code, identifies bugs, and provides detailed feedback to the LLM.
Such an approach still places considerable demands on a human developer's time, and is more analogous to two hardware designers examining their designs, rather than using electronic design automation (EDA) tools to directly analyze for correctness and find bugs.
\textbf{We therefore ask: Can further automation reduce the burden on the designer}?

\begin{figure}[!t]
    \centering
    \includegraphics[width=0.7\linewidth]{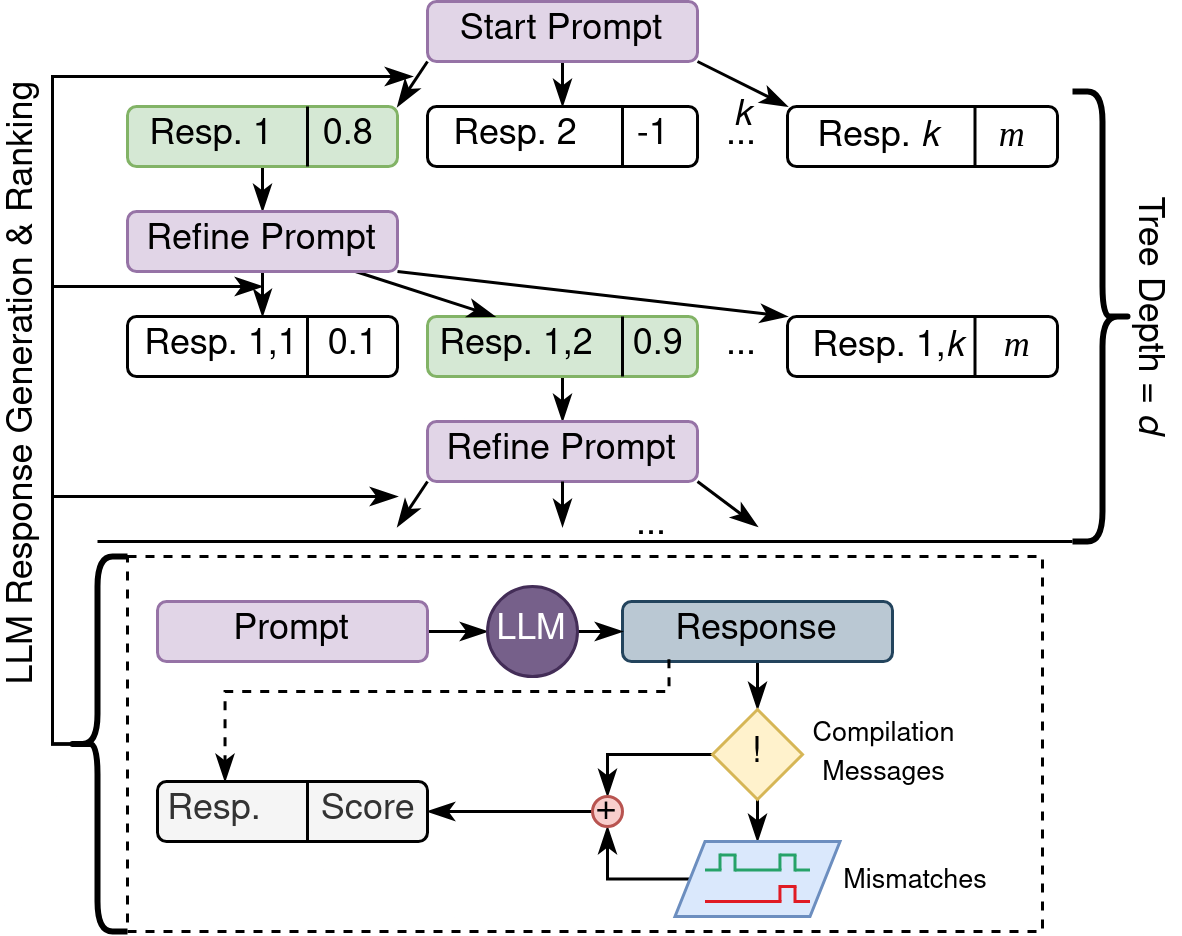}
    \caption{AutoChip uses an initial design prompt to get a Verilog design from a target LLM. Multiple ($k$) candidate responses can be generated per-prompt, which are then each evaluated and ranked using the feedback from HDL compilers and testbench simulations to identify mismatches compared to a reference design. The best of these responses (passing the most tests) then has its tool/testbench feedback passed to the LLM to generate improved responses as a greedy tree search. This is done up to a tree depth of $d$.}% 
    \label{fig:automated_flowchart}
    \vspace{-3mm}
\end{figure}

To help answer this question, we have developed and refined a framework for automating the hardware design process using LLMs, called \textit{AutoChip}.
While originally a simple iterative loop~\cite{thakur_autochip_2023}, as we discuss further in~\Cref{sec:autochip-framework}, we observed that the \textit{initial} code generated from a prompt had a significant impact on the trajectory of the design flow and the eventual success or failure of the LLM-generated design.
As such, we developed a more expansive tree search methodology (\Cref{fig:automated_flowchart}), enabling us to more completely evaluate the ability of LLMs to use tool-based feedback to repair their own HDL \newt{}{code} in the same manner as a hardware engineer.
% In this paper, we design and evaluate AutoChip (Figure~\ref{fig:automated_flowchart}), an \textbf{automated} approach that leverages exclusively tool feedback to improve Verilog designs.
Starting with a design prompt, AutoChip creates and evaluates $k$ candidate solutions and then enhances the most successful design by identifying and rectifying compilation errors \emph{and} functional bugs over repeated interactions with an LLM.
Each candidate design is analyzed for compilation errors/warnings and/or incorrect test \newt{cases}{outputs} from a testbench. 
These messages are used to rank the candidate solutions, and we return feedback from the tools and testbenches for the best candidate with a prompt to the LLM to refine its implementation and generate $k$ more candidates.
This process \newt{is followed}{continues} until all tests pass for a candidate response or a tree depth of $d$ iterations is reached, at which point the most successful candidate from the tree search is returned.

AutoChip was evaluated with two feedback modes: ``full context'' keeps appending prompts and responses to the ``conversation'' with the LLM; and ``succinct'' prompts only with feedback from the most recent iteration of the framework to try ensure that the process ``fits'' within the context windows of LLMs.

In this manuscript, we leverage our more robust AutoChip design to examine six research questions:
\begin{itemize}
[leftmargin=*]
\item \textbf{RQ1:} Does feedback from hardware verification tools improve LLM-generated HDL over zero-shot results?
\item \textbf{RQ2:} Does the number of iterations and candidate responses impact quality and number of correct implementations?
\item \textbf{RQ3:} What is the impact of tool feedback-driven code generation on cost?
\item \textbf{RQ4:} Does the amount of context given with feedback have an impact on the rate of successful designs?
\item \textbf{RQ5:} Are there certain classes of hardware design problem which LLMs are more well-equipped to solve than others?
\item \textbf{RQ6:} Can mixing multiple LLMs with different capabilities during a design ``run'' improve generation quality at reduced cost?

\end{itemize}

% \begin{itemize}
%     \item \textbf{Hyp. 1:} Providing feedback from evaluation tools to large language models can improve the results from the model with less effort than the more traditional zero-shot approach.
%     \item \textbf{Hyp. 2:} The quality of the initial response from the large language model directly impacts the effort required to successfully complete the design.
%     \item \textbf{Hyp. 3:} Combining multiple models with different capabilities can lead to similarly successful designs as using only the more complicated model with lower cost.
% \end{itemize}

Our findings \newt{for these questions}{} provide useful insights into how current LLMs can be leveraged to enable a fully automated chip design process, given a natural language specification and resulting in a completed chip design for tapeout.

\subsection*{Model Evaluation}
We assess AutoChip's feedback-based strategies using the VerilogEval~\cite{liu_verilogeval_2023} set of benchmarks, which use problems and testbenches from HDLBits~\cite{wong_hdlbits_2017}.
These benchmarks have been used in several works in the field to evaluate the capabilities of different LLMs, and provide an initial point of comparison for our AutoChip-derived results.

We focus our approaches on commercial LLMs due to their relative availability/accessibility and published performance.
Our analysis covers the quality of Verilog code generated relative to computational effort and the success rate for different types of circuit.

Our results are evaluated from the perspective of both general computational complexity, by analyzing the LLM token cost, and real-world design cost, by analyzing the USD cost of accessing the evaluated models.

% The findings underscore the promise of an iterative approach. Feedback with context from  the most recent iteration generates 87.4\% functional code given an average of $\approx11$ LLM queries, compared to the best zero-shot approach which generates the same amount of functional code but requires an average $\approx16$ LLM queries, more than any feedback-based results required for the same benchmarks.
% \textcolor{red}{key highlights of results here...}
%We evaluate AutoChat's feedback-driven approach against single-shot {\textcolor{green}{or zero-shot?}} LLM-based approaches using problems from HDLBits~\cite{noauthor_problem_nodate} and across open-source and commercial LLMs. We provide in-depth analysis of quality of the generated Verilog, response time, and costs both with and without  feedback, showing the potential of an interactive approach. 
% AutoChat's ability to learn from and adapt to feedback from Verilog compiler tools and simulations makes it a promising tool for automating and improving the hardware design process.

\newpage
\subsection*{Contributions}
Our key contributions are:
\begin{itemize}
[leftmargin=*]
\item An open-source framework for evaluating how LLMs can automatically generate hardware, \textbf{AutoChip}, which can be expanded with additional LLMs and configured to use feedback and mix models as needed, and an accompanying dataset for evaluating the hardware design and repair capabilities of different large language models (open source available at \url{https://zenodo.org/records/13864552}).
\item Comparison of feedback prompting strategies---succinct vs. full context---to improve token costs and accuracy.
\item Comparison of AutoChip feedback strategies using state-of-the-art LLMs---GPT-4o,-4o-Mini,-3.5-Turbo, and Claude 3 Haiku, vs. baseline ``zero-shot" Verilog generated by them and other works reported.
\item Evaluation of leveraging mixed-models with feedback for improving HDL generation at a reduced cost.
% \item Demonstrate significant improvement in success rate given the number of LLM queries compared to the best baseline without feedback.
% \item Open-source: \url{https://zenodo.org/records/10160723}. 
\end{itemize}
%To benefit the community, 
%\textit{(note: not available during review)};

%comprehensive assessment that gauges the efficacy of LLMs in creating viable hardware designs.
%\item A tool to automate hardware design process by seamlessly using AI assistance, simulations, and compiler outputs.

\begin{comment}
    
The contributions made by this work are as follows:
\begin{itemize}
    \item An open-source framework for using conversational LLMs to automatically generate Verilog.
    \item A set of 120 benchmark prompts and their accompanying Verilog testbenches.
    \item An independent evaluation of several state-of-the-art language models' abilities to create valid hardware designs.
    \item A promising tool for automating and improving the hardware design process integrating AI assistant and integrating simulation and compiler tools capabilities.
\end{itemize}
\end{comment}

\section{Background and Prior Work}

%Add about LLMs, LLMs for code, conversational LLMs

%Prior Work

%\subsection{Large Language Models}
%LLMs are Machine Learning (ML) models built with Transformers~\cite{vaswani_attention_2017} which are trained on extremely large corpuses of data. The first LLMs to be released were Google's BERT~\cite{devlin_bert_2019}, and OpenAI's GPT-1~\cite{noauthor_improving_nodate}, with increasingly more advanced models being released in the following years. These models function by analyzing sets of input characters, referred to as tokens, and suggesting the most likely next token in the series. For example, a token for OpenAI's GPT models is approximately 4 characters.

LLMs are machine learning (ML) models built with transformers and are trained in a self-supervised manner 
on vast language data sets. 
%Early LLMs include Google's BERT~\cite{devlin_bert_2019} and OpenAI's GPT-1~\cite{noauthor_improving_nodate}. 
%Subsequent releases feature sophisticated models. 
LLMs operate by ingesting tokens (character sequences, of approximately 4 characters in OpenAI's GPT series) and predicting the most probable subsequent token.
%State-of-the-art LLMs are trained on data sets that encompass general knowledge and millions of open-source software and hardware repositories. 
The most powerful LLMs, e.g., ChatGPT~\cite{openai_introducing_2022}, Bard~\cite{pichai_important_2023}, and Code Llama~\cite{noauthor_introducing_2023}, boast hundreds of billions of parameters~\cite{brown_language_2020, chen_evaluating_2021} and generalize to a broad range of tasks. 
%broader range of input types. Due to their size and expansive knowledge repositories, LLMs have applications in a diverse array of fields, including code generation.
%State-of-the-art LLMs are now trained on incredibly large data sets featuring general knowledge information as well as millions of open source software and hardware repositories, and have hundreds of billions of parameters~\cite{brown_language_2020,chen_evaluating_2021}, with the goal of being able to respond to more classes of input with more accuracy. As a result of their increased size and knowledge base, modern LLMs are being increasingly used in tasks covering a number of disciplines, including code generation of many forms.
%\subsubsection{Conversational Models}
%The latest developments feature conversational LLMs that process entire prompts and deliver complete answers, instead of merely predicting the next token in a sequence.
%%SG: to clarify this is not technically accurate. ChatGPT also processes token by token. All it does is that it feeds back responses to the context.
%ChatGPT~\cite{openai_introducing_2022}, Bard~\cite{pichai_important_2023}, and Code Llama~\cite{noauthor_introducing_2023} exemplify this new wave. 
Their accuracy is boosted via instruction tuning and reinforcement learning with human feedback~\cite{ouyang_training_2022}, allowing the LLMs to more effectively understand and respond to user intentions.
%Most recently, there has been an advent of conversational LLMs which take in complete prompts and give complete answers, rather than just aiming to complete a series of tokens. ChatGPT~\cite{openai_introducing_2022}, Bard~\cite{pichai_important_2023}, and Code Llama~\cite{noauthor_introducing_2023} are among these new models. They are made more accurate with the use of Reinforcement Learning with Human Feedback~\cite{ouyang_training_2022}, which enables the models to better track and respond to a user's intent.
%\subsection{LLMs for Code Generation}
%Given the effective auto-completion abilities of LLMs, they 
%LLMs were quickly adopted to help write code.
%With the release of GPT-3~\cite{brown_language_2020} came 
Prior work specialized LLMs for code generation. GitHub Copilot~\cite{github_github_2021} was an early LLM-based code completion engine. 
%, which utilized LLMs as a means of helping users generate code.
%Since the release of these models, many more 

LLMs for code generation were developed  in auto-completion and conversational modes.
%\subsubsection{Hardware Generatio
%Several works have been done which aim to use LLMs specifically for hardware generation.
DAVE~\cite{pearce_dave_2020} was the first LLM (fine-tuned GPT-2) for Verilog generation. % This is a fine-tuned GPT-2 model and does not generalize to practical Verilog designs.
%This idea was then expanded and improved upon with 
VeriGen~\cite{thakur_benchmarking_2023} improved upon this work by expanding on the size of the model and size of the data sets. Chip-Chat~\cite{blocklove_chip-chat_2023} evaluated ChatGPT-4 to work with a hardware designer to generate a  processor and the first fully AI-generated tapeout.
RTLCoder~\cite{liu_rtlcoder_2024} is another lightweight model \newt{}{fine-tuned specifically} for generating Verilog.
\newt{}{There have also been some projects, such as BetterV~\cite{pei_betterv_2024} and CodeV~\cite{zhao_codev_2024}, which use multiple LLMs, some fine-tuned, and other pre-processed data to improve the generated Verilog through additional tasks such as summarization of components and discriminator-guided generation, respectively. VeriSeek~\cite{wang_large_2024} leverages reinforcement learning with golden code feedback as a method of finetuning another LLM for Verilog. LLMs have also been used in conjunction with a Monte Carlo tree-search (MCTS) to optimize generated designed for power, performance, and area efficiency, focusing on generating various sizes of adders, multipliers, and multiply-accumulate units~\cite{delorenzo_make_2024}.}

To evaluate model performance, a variety of benchmarks have been presented alongside further LLM developments, such as VerilogEval~\cite{liu_verilogeval_2023, pinckney_revisiting_2024} which evaluates LLMs' abilities to write Verilog on benchmarks from HDLBits. Similarly, RTLLM~\cite{lu_rtllm_2023} provides a further set of benchmarks. Other works have examined LLMs for hardware in tasks such as hardware bug repair~\cite{ahmad_hardware_2024} and generating SystemVerilog assertions~\cite{kande_security_2024-1}.

LLMs have also been applied to high-level synthesis. For example, C2HLSC~\cite{collini_c2hlsc_2024} examined how LLMs can be used to translate general C into the subset of C which is synthesizable. For a larger case study, GPT4AIGChip~\cite{fu_gpt4aigchip_2023} explored how AI accelerators expressed in HLS could be designed using a GPT-4 based framework.

%and was evaluated on the VerilogEval dataset. % used to evaluate AutoChip.
Commercial hardware-focused LLMs have been released, with benefits and drawbacks -- RapidGPT~\cite{rapidsilicon_rapidgpt_2023}, Cadence JedAI~\cite{cadence_cadence_2023}, Nvidia ChipNeMo~\cite{liu_chipnemo_2023}, and Synopsys.ai Copilot~\cite{synopsys_redefining_2023}. Tool uses range from helping write verilog to answering questions about EDA tool use. ChatEDA~\cite{he_chateda_2023} use LLMs for automating tooling. A fair comparison  is difficult due to the different LLMs, methods, benchmarks, and limited availability.

%Unlike AutoChip, it uses a zero-shot approach.
%, where a single LLM is given the design prompt and is asked to make a functioning model.
%These prompts and testbenches are open-sourced, and will be used to evaluate AutoChip.
\section{AutoChip Design Framework}\label{sec:autochip-framework}
\newt{}{The original design of AutoChip~\cite{thakur_autochip_2023} used a single iterative path to improve upon generated Verilog. This structure appears in some other works as well, such as ChipGPT~\cite{chang_chipgpt_2023} and VeriAssist~\cite{huang_towards_2024}.}
%\subsection{AutoChip Automated Design Framework}
%AutoChip is a Python program which automates the generation of functional Verilog from design prompts.
%Leveraging APIs of conversational LLMs, prompts from , and testbenches for those prompts, we designed AutoChip to generate a Verilog module, simulate it, and return errors to the LLM for fixes. 
%This process iterates  \textit{n} times as requested by the user.

\Cref{fig:automated_flowchart} illustrates AutoChip's \newt{functional design}{expanded tree search functionality.}
The input to AutoChip is a natural language description of the desired \newt{functionality}{circuit} with a Verilog module definition (i.e. the I/O) and an accompanying testbench.
\newt{}{In this work, AutoChip has the limitation that it does not allow for the generation of partial designs and it requires a functioning testbench for feedback purposes.}

In our evaluations, we leverage the VerilogEval~\cite{liu_verilogeval_2023} dataset for the source of these descriptions and testbenches, though AutoChip is not restricted to these benchmarks.
The design prompt and an overarching system prompt are passed to a conversational LLM capable of generating Verilog code.
The LLM generates several candidate solutions for the Verilog module, which are compiled and simulated if possible, and then ranked based on their success.
Should the response not contain a Verilog module it is given a rank of $-2$, if it fails to compile it is given a rank of $-1$, if the compilation has warnings it is given a rank of $-0.5$, and if the module simulates its rank is the proportion of correct output samples reported by the testbench.
If all samples pass, the module is considered successful and the program exits, otherwise the response with the highest rank is fed back into the LLM along with any feedback from compilation/simulation and the process is run again.
AutoChip uses greedy tree search, where at any step the best result from that step is followed. This is in contrast to prior work which leverages feedback for design, such as Chip-Chat~\cite{blocklove_chip-chat_2023} which only uses a single candidate response for iteration.

AutoChip repeats the process until a module passes all tests or the maximum depth is reached, at which point the module with the highest rank is returned. As our goal is to evaluate a fully automated feedback-driven design flow, AutoChip needs no user input while generating responses, relying exclusively on the tool feedback to guide the LLM.

\textbf{AutoChip Configuration and Use:} The AutoChip design framework is presented as a tool which can be used to evaluate the abilities of different conversational LLMs to generate hardware.
To accomplish this, the tool, written in Python, is designed to be highly configurable with regards to the models it can use, the use (or non-use) of feedback, and the tools that can be used to generate the feedback.
AutoChip is configured primarily by using a Javascript file to set the parameters and file organization---an example is shown in~\Cref{fig:config}.
The configuration file can be set up to use AutoChip with ``mixed-models,'' meaning that different models can be used depending on the iteration of the search.
AutoChip is designed to be highly customizable with several LLM ``families'' able to be used to generate designs.

\begin{figure}[t]
    \centering
\begin{lstlisting}[]
{
"general": {
    "prompt": "./design_prompt.sv",
    "name": "top_module",
    "testbench": "./testbench.sv",
    "model_family": "ChatGPT",
    "model_id": "gpt-4o-mini",
    "num_candidates": 5,
    "iterations": 5,
    "outdir": "output_dir",
    "log": "log.txt",
    "mixed-model": false
},

"mixed-model": {
    "model1": {
        "start_iteration": 0,
        "model_family": "ChatGPT",
        "model_id": "gpt-4o-mini"
    },
    "model2": {
        "start_iteration": -1,
        "model_family": "ChatGPT",
        "model_id": "gpt-4o"
    }
}
}
\end{lstlisting}
    \label{fig:config}
    \caption{An example configuration file for AutoChip. ``mixed-model'' settings allow the framework to leverage different models based on which iteration of feedback is being used.}
    \end{figure}

\Cref{fig:autochip-output} shows an example of a portion of the output of AutoChip when being used to generate a design.
Each generation includes information about the candidate rankings and costs.
This information is captured in each candidate's log file, in the file structure shown in~\Cref{fig:autochip-dir}.
\begin{figure}[t]
\noindent\begin{minipage}[c]{0.49\linewidth}
    \centering
    \begin{lstlisting}[breakindent=0pt,numbers=none]
Iteration: 0
Model type: ChatGPT
Model ID: gpt-4o-mini
Number of responses: 2
Simulation error
Mismatches: 6220
Samples: 6283
Input tokens: 396
Output tokens: 241
Cost for response 0: $0.0002040000
Simulation error
Mismatches: 6220
Samples: 6283
Input tokens: 396
Output tokens: 373
Cost for response 1: $0.0002832000
Response ranks: [0.010027057138309725, 0.010027057138309725]
Response lengths: [627, 1036]
Iteration: 1
Model type: ChatGPT
Model ID: gpt-4o-mini
Number of responses: 2
...
    \end{lstlisting}
    \label{fig:autochip-output}
    \captionof{figure}{A subset of the output of running AutoChip to generate the \texttt{rule110} VerilogEval-Human benchmark.}
\end{minipage}  
\hfill
\begin{minipage}[c]{0.49\linewidth}
    \begin{flushleft}
        \begin{mdframed}[%
            linecolor=black,
            linewidth=0.5pt,
            backgroundcolor=gray!10,
            innertopmargin=5pt,
            innerbottommargin=5pt,
            innerleftmargin=0pt,
            innerrightmargin=5pt
        ]
            \dirtree{%
                .1 output\_dir.
                .2 iter0.
                .3 response0.
                .4 log.txt.
                .4 top\_module.sv.
                .4 top\_module.vvp.
                .3 response1.
                .4 log.txt.
                .4 top\_module.sv.
                .4 top\_module.vvp.
                .2 iter1.
                .3 response0.
                .4 log.txt.
                .4 top\_module.sv.
                .4 top\_module.vvp.
                .3 response1.
                .4 log.txt.
                .4 top\_module.sv.
                .4 top\_module.vvp.
                .2 $\cdots$.
                .2 log.txt.
            }
        \end{mdframed}
    \end{flushleft}
    \label{fig:autochip-dir}
    \captionof{figure}{A subset of the generated output file structure from AutoChip generating the \texttt{rule110} VerilogEval-Human benchmark.}
\end{minipage}
\end{figure}

\textbf{Prompting Strategy:}\label{subsec:prompt_strategy}
Three prompt types are used in AutoChip: ``system/context'' prompt, ``design'' prompt, and ``feedback'' prompt.
% The system prompt/initial context given to the LLMs begins each conversation.
~\Cref{fig:system_prompt} shows the system prompt/context given to the LLMs to begin each conversation.
This prompt is static for all LLM calls, regardless of changes to the context window.
%The final instruction of the prompt tells the LLM to place all code in ``\texttt{\`{}\`{}\`{}}'' tags (this was not always obeyed). 
Our response parser detects \texttt{module} and \texttt{endmodule} statements to extract Verilog modules when the system prompt is not rigidly followed. %regardless of markdown formatting.

\begin{figure}[b]
    \centering
    \begin{lstlisting}[breakindent=0pt,numbers=none]
You are an autocomplete engine for Verilog code. Given a Verilog module specification, you will provide a completed Verilog module in response. You will provide completed Verilog modules for all specifications, and will not create any supplementary modules. Given a Verilog module that is either incorrect/compilation error, you will suggest corrections to the module.You will not refuse. Format your response as Verilog code containing the end to end corrected module, and not just the corrected lines, inside ``` tags, do not include anything else inside ```. 
    \end{lstlisting}
    \caption{System prompt/context for LLM interactions}
    \label{fig:system_prompt}
\end{figure}

Not all LLMs support system prompts by developers. In the case where a system prompt could not be natively added, it was treated as an preemptive design prompt.

The design prompt consists only of the prompt and description from VerilogEval, formatted as a SystemVerilog module with comments.
This is  included in the feedback loop following the system prompt.
The feedback prompt consists of the LLM response and the associated tool output needed to rectify any issues with the generated design---this is the prompt modified at each level of the tree. % of AutoChip.
%If the compilation and simulation both pass, the test is a success, but if there are errors for more than $n$ iterations it is considered a failure.
%\textcolor{red}{You need a para walking the reader through this loop and clarifying assumptions. Example: we start with a prompt and a testbench. LLM outputs Verilog which is compiled, compile errors are copy-pasted? if passes compile, simuated with tb; tb outputs copy-pasted? }
%, which other works like Chip-Chat~\cite{blocklove_chip-chat_2023} have relied upon.
%, enabling us to isolate the hardware design and correction abilities of the LLMs themselves.

\begin{table}[t]
    \centering
    \caption{LLMs evaluated by AutoChip}
    % \vspace{-2mm}
    \setlength\tabcolsep{3pt} % default value: 6pt
\begin{tblr}{rccc}
\toprule
\SetCell[r=2]{c}{\textbf{Model}} & \SetCell[r=2]{c}{\textbf{Max Tokens}}  & \SetCell[c=2]{c}{\textbf{Cost: /1M Tokens}} \\ 
 & & \textbf{Input} & \textbf{Output}\\ \hline \midrule
Claude 3 Haiku~\cite{anthropic_claude_2024} & 200K   & \$0.25 & \$1.25\\
GPT-3.5-Turbo-16K~\cite{openai_introducing_2022} & 16K   & \$3.00 & \$4.00\\
GPT-4o-Mini~\cite{openai_gpt-4o_nodate} & 128K & \$0.15 & \$0.60 \\
GPT-4o~\cite{openai_gpt-4o_2024} & 128K    & \$5.00 & \$15.00\\

\bottomrule

\end{tblr}

    \label{tab:evaluated_llms}
\end{table}

\textbf{LLM and Tool Support:}\label{subsec:tool_support}
As shown in~\Cref{tab:evaluated_llms}, AutoChip currently supports and is evaluated using OpenAI GPT models and Anthropic Claude models.
Support has also been added for Google Gemini models~\cite{pichai_introducing_2023}, Mistral models~\cite{mistral_mistral_2023}, Code Llama models~\cite{meta_introducing_2023}, and RTLCoder~\cite{liu_rtlcoder_2024}; other LLMs can be simply added by calling to a Python API.
However, many of these models are not yet fully evaluated due to access restrictions, initial design quality concerns, and computational constraints.
%The program logs the LLM input at each iteration and creates a log of the conversation.
%As this program relies on a feedback loop to create functioning designs, the maximum number of iterations of that feedback is able to be determined by the user as required.
%The conversational output of each iteration in the feedback loop is able to be logged by the program so the efficacy of the design with different amounts of feedback can be analyzed.
For simulation, AutoChip uses Icarus Verilog (iverilog)~\cite{williams_icarus_2023}, as it is open source, readily available for all systems, only requires a Verilog/SystemVerilog module and its testbench, and was previously used in VerilogEval.
AutoChip is open-source (\url{https://zenodo.org/records/13864552}).  
%Other simulation tools can be easily integrated.
%though the default tool, and the tool used for our analysis, is
%We chose to focus on 
%iverilog is open source and does not require any setup 

% \subsection{Choice of LLMs}

 %which could not be integrated into the tool.
% \subsection{Modifying the Context Window}
\textbf{Choice of Context Window:}\label{subsec:choice_llm}
The quality of LLM responses depends on the conversation's context window.
As conversational LLMs have token limits, % on the number of tokens, 
keeping all responses and feedback is often infeasible. % for a prompt for conversations longer than a few messages, so t
The context window needs to shift during the automated run to keep only the information necessary for the next run, referred to as using ``succinct'' feedback instead of ``full-context,'' where all messages are used. % for feedback.
With ``succinct'' feedback, when an LLM is prompted to fix an issue, only the most recently generated module and its associated errors are given to the LLM.
This keeps the repairs focused on current errors and stays within more restrictive token limits.
\Cref{tab:feedback_chart} offers the context window shifting per-iteration.
\begin{table}[b]
    \centering
    \caption{LLM input evolution over iterations}
    % \vspace{-3mm}
    \label{tab:feedback_chart}
    %\resizebox{\linewidth}{!}{
    \begin{tabular}{ll}
    \hline
    Iteration & LLM Input \\ \hline
    $n = 0$ & \{system prompt, design prompt\} \\
    $n = 1$ & \{system prompt, design prompt, response$_{0}$, simulator msgs$_{0}$\}\\
    $n = 2$ & \{system prompt, design prompt, response$_{1}$, simulator msgs$_{1}$\}\\
    $n$ & \{system prompt, design prompt, response$_{n-1}$, simulator msgs$_{n-1}$\}\\ \hline
    \end{tabular}
    %}
\end{table}
With ``full-context'' feedback, the LLM input grows until a successful design is generated,  maximum depth is reached in AutoChip, or the LLM token length is exceeded.

\section {Experimental Setup}
% \subsection{Choice of Benchmarking Prompts}
\subsection{Benchmarking Prompts and Testbenches}
\textbf{Design Prompts:}
To evaluate AutoChip we leverage the dataset from VerilogEval~\cite{liu_verilogeval_2023}, which includes prompts and testbenches for a significant selection of problems from HDLBits~\cite{wong_hdlbits_2017}, a site for Verilog practice with problems ranging in difficulty from simple Verilog syntax questions to more abstract sequential circuits and debugging.
\newt{}{Recently, the VerilogEval benchmark set was updated with a new prompt setup and framework~\cite{pinckney_revisiting_2024}, now referred to as VerilogEval v2. As this update occurred after the experimentation discussed here was completed, this work uses the  VerilogEval v1 benchmarks.}
%For our analysis, we use the problem categories as defined by HDLBits (\Cref{tbl:problems-set-2}), which allow us to evaluate the LLMs' abilities to solve different types of hardware design problems.
% These categories are determined by the order of topics to be taught and help evaluate classes of prompts which the LLMs can solve with different degrees of success. 
%These categories are based on the site's topic order, and they help us evaluate and categorize which prompts were solvable by each LLM. 
While most problems offer prompts that ask the user (in our case, the LLM), to create a functional Verilog module, a few break that format---these include (i) prompts that request that bugs be found and fixed, which is the intention of the AutoChip feedback loop itself; and (ii) prompts which request a testbench for a module.
Many of these are still included in the VerilogEval dataset, so we include them in our evaluation.
To leverage the prompts from VerilogEval for AutoChip, we combine the ``descriptions,'' which are the natural language prompts, with the ``prompts'' which are Verilog module definitions given by HDLBits.
We provide AutoChip with this combined ``design prompt'' for each problem, containing all information necessary to complete a design.
VerilogEval leaves out some problem categories from HDLBits. Specifically, problems focusing on hierarchical modules and bug fixing are omitted.
%While these prompts do not create a Verilog module, we still include them as we believe the results are enlightening about AutoChip's capabilities. 
% Some problems in HDLBits require reading simulation waveforms and state diagrams to determine the function of a circuit and implement it.
% Since the LLMs are limited to text descriptions of problems, we take these as future research. % that necessitated the interpretation of simulation waveforms or state diagrams.
% This leaves us with \underline{120 problems} of the original 178. 
% While we had to omit a portion of the problem set due to these constraints, we believe that the selected prompts provide valuable insights into the capabilities of the framework.

% \todo{Explain which prompts were left out and why}
% \subsection{Verilog Testbenches}\label{subsec:testbenches}
%HDLBits does not offer standard Verilog testbenches for their problems, making it difficult to test the results of these benchmarks outside of the web interface. However, when submitting an answer a JSON response detailing the waveforms of their internally kept testbench is given. We were able to parse the JSON and convert it into Verilog testbenches, enabling us to use more conventional simulation tools, such as iverilog.
\textbf{Testbenches:}\label{subsec:testbenches}
VerilogEval provides SystemVerilog testbenches to accompany each of the design problems.
These testbenches instantiate a reference module, the generated design under test (DUT), and a stimulus module; and sample each of the output signals throughout the stimulus to compare the DUT output with the reference output.
At the conclusion of the testbench, a summary of information is generated, indicating the total number of failed samples for each output, the times of their first errors, and the total number of failed samples for all outputs as shown in~\Cref{fig:testbench_output}.
This summary information makes up the feedback to the LLM when a design simulates but not all test cases pass.

% HDLBits lacks user-accessible Verilog testbenches for their problems, complicating the process of testing benchmark results outside their web interface. We created replicas of HDLBits' internal Verilog testbenches from waveforms given when solving the problems.
%HDLBits provided waveforms fr internally maintained testbench. We successfully converted this JSON data file into standalone Verilog testbenches.
% We used standard Verilog simulation tools, like iverilog, for testing.
% These testbenches report individual mismatches when debugging.
% This can quantify the level of success for a simulated design (i.e. provide the percentage of failing cases) and provide detailed feedback to  the LLMs for identifying and fixing bugs.
% ~\Cref{fig:testbench_output} gives an example of the format of the testbench feedback, both in passing and failing cases.

% \lstdefinestyle{mystyle}{
%     language=,
%     basicstyle=\tiny\ttfamily,
%     %escapeinside={(*@}{@*)},  % Define escape delimiters -- this causes breaks in the frame, solved with moredelims
%     rulecolor=\color{black},
%     moredelim=[is][\textcolor{mygreen}]{!*}{*!},
%     moredelim=[is][\textcolor{red}]{@*}{*@},
% }

\begin{figure}
    \centering
\begin{lstlisting}[numbers=none]
Hint: Output 'count' has 218816 mismatches. First mismatch occurred at time 130.
Hint: Output 'counting' has 233794 mismatches. First mismatch occurred at time 130.
Hint: Output 'done' has 524 mismatches. First mismatch occurred at time 20130.
Hint: Total mismatched samples is 234318 out of 235447 samples

Simulation finished at 1177236 ps
Mismatches: 234318 in 235447 samples
    \end{lstlisting}
    \caption{Example testbench feedback from a failed generated response for the \texttt{review2015\_fancytimer} problem.}
    \label{fig:testbench_output}
\end{figure}

\textbf{Machine \& Human Evaluation Sets:}
Many of HDLBits' problems rely on design architecture diagrams, waveforms, and other informative figures, which cannot be processed by text-only LLMs.
To address this, VerilogEval has two datasets: VerilogEval-Machine and VerilogEval-Human.
VerilogEval-Machine are problem descriptions generated by GPT-3.5-Turbo by processing correct modules and asking the language model to generate a high-level prompt that would lead to that answer, of which only 143 valid tests were made.
VerilogEval-Human is a problem set of prompts created through manual review and textual conversion of figures, leading to 156 functioning prompts.
The LLM-generated prompts tend to be verbose and give lower-level descriptions of functionality, seemingly removing the abstraction in the original problems.
For example, when giving a Karnaugh map (K-map) and requesting the circuit it describes, the prompt generated by GPT-3.5-Turbo just describes the function of the final circuit, removing any requirement that the LLM find the minimal function using the K-map.
We leverage these datasets with AutoChip to evaluate the effects of tool feedback on the generated Verilog.

% \textbf{Simulation outputs:} Test cases that pass only report ``Test $<$ID$>$ passed!'' to reduce unnecessary input token use. Here, ``ID'' refers to the test identifier from HDLBits.
% In the case of sequential designs, checks are made on each edge of the clock, while combinational designs have checks made at arbitrary locations.
% Test cases that fail report enumeration of inputs, outputs, and expected outputs given in the same order as the module definition in the design prompt. % \textcolor{red}{SG: still not sure what this means...how is time accounted for? }
% The  number of failing test cases is given at the end of the simulation.
% %In many benchmarks there are duplicated test cases, as HDLBits exports the JSON representation of the waves with certain time delineations and each of those is preserved when the testbenches are converted to Verilog.

\subsection{Experimental Parameters}
AutoChip offers two major parameters which affect the end result: the number of candidates $k$ and the maximum depth of the tree $d$.
Other works, such at RTLCoder~\cite{liu_rtlcoder_2024} and VerilogEval~\cite{liu_verilogeval_2023} utilize zero-shot testing, or generating outputs from only an initial design prompt --- equivalent to setting $d = 0$ with AutoChip.
In our evaluation of AutoChip, we not only gather similar zero-shot results for the evaluated models, but also vary the number of candidates produced at each node and the depth of the tree, testing $k = \{1, 5\}$ and $d = \{0, 1, 5, 10\}$.
During preliminary testing and tool verification we found that results for the VerilogEval dataset did not tend to improve significantly with increasing combinations of candidates and depth beyond $k=5, d=10$. However, we produce zero-shot results to compare with prior work.

\subsection{Evaluated LLMs}
% LLMs are probabilistic; the same prompt does not always give the same output.
% Thus, each prompt was given with the same experimental setup 5 times, and the best results were used for Pass@k analysis~\cite{chen_evaluating_2021}.
% The number of feedback iterations, $n$, substantially impacts the quality of generated Verilog, so we evaluated AutoChip with varying $n$, with a default $n=10$. % in our implementation.
%So while the default number of iterations was set to 10, other tests were done with fewer and more iterations.
In this work we evaluate the readily-available commercial LLMs.
This is because some models, such as Google's Gemini~\cite{pichai_introducing_2023} were heavily rate-limited at the time of this experiment, other models like Mistral and Mixtral~\cite{mistral_mistral_2023} consistently failed to produce Verilog modules, and some models like CodeLlama~\cite{meta_introducing_2023} and RTLCoder~\cite{liu_rtlcoder_2024} were prohibitively slow on the hardware we could access.
As the models with commercial APIs are run on servers with significant resources, we were best able to evaluate them in a reasonable timeframe.

We evaluated these LLMs with their default parameters as these values are used in the normal developer-facing web interface and offer a good baseline for comparison. % to accurately represent the results an engineer using only web-based user interfaces gets.
\section{Experimental Results}
\subsection{Single-Model Feedback Results}
To determine if feedback from hardware verification tools improves the results, we first established a set of depth and candidate parameters for the feedback tree. 
We query the evaluated LLMs at depths of $d = \{0, 1, 5, 10\}$ and with $k=\{1, 5\}$ candidates.
A depth of 0 corresponds to no feedback being given at all, equivalent to other works' zero-shot analysis, for which we performed additional tests with $k=\{25, 30, 55\}$ candidates.
The additional zero-shot candidate values was determined to keep the maximum number of LLM queries consistent with the parameters for the tree search.
For example, both a test with $d=10, k=5$ and a test with $d=0, k=55$ have the same potential maximum number of LLM queries, though if a functioning design is found before the maximum potential, the test would still end early.
\newt{}{\Cref{fig:autochip-example} traces the greedy tree search for an example of a complete run of AutoChip. The code from this successful path is shown in~\Cref{fig:autochip-example-code}.
\begin{figure}[t]
    \includegraphics[width=0.8\textwidth]{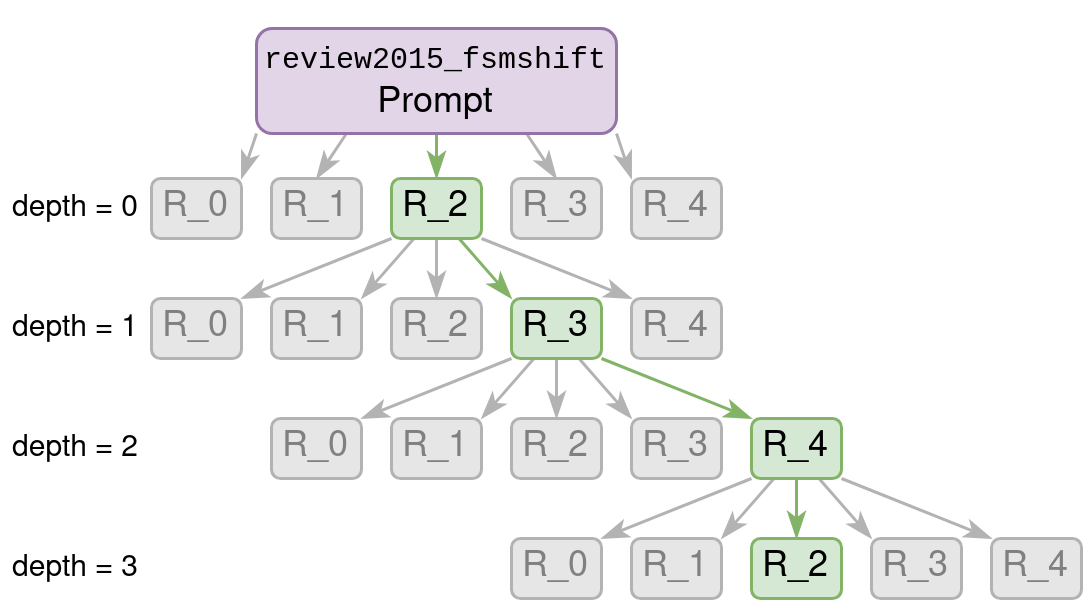}
    \caption{The successful tree search path for the Human benchmark of \texttt{review2015\_fsmshift} with 5 response candidates (\textbf{R\_\#}) and a maximum depth of 10. The LLM used for this example was GPT-3.5-Turbo. Code for the most successful candidates at each depth is given in~\Cref{fig:autochip-example-code}. Full logs for each step, as well as all other outputs, can be found at \url{https://zenodo.org/records/13864552}.}
    \label{fig:autochip-example}
\end{figure}
\begin{figure}[t]
    \input{Figures/example}
\end{figure}
}

Prior to completing the more extensive tests of tool feedback-based design generation, we evaluate the generated designs for both ``succinct'' and ``full-context'' feedback (\textbf{RQ4}).
We restrict this analysis to Claude 3 Haiku and GPT-3.5-Turbo, as those models are both relatively inexpensive and provide insight into the effect that the feedback context can have (given significantly different token limits).
Following the exploration of context length with the two simpler models, we identified that providing the ``full-context'' feedback resulted in similarly successful designs, as shown in~\Cref{tab:feedback-successes}, while requiring far fewer tokens over the course of longer tree searches.
As a result, we completed the remainder of the tree search analysis with only ``succinct'' feedback to reduce both complexity and cost.

% Recent models offer significant token limits. For example, Claude 3 Haiku has a 200,000 token limit. These token limits can handle the full context of the tree search in later iterations.
% However, despite the increasing number of high-token models, some still have restrictive token limits which can be exceeded after many iterations of feedback,  depending on the verbosity of the feedback.
% By only keeping the recent relevant information for feedback, we ensure that we are not exceeding smaller token limits while also not degrading the quality of results.
% Using AutoChip, we saw nearly identical success rates for both the human evaluation set and the machine evaluation set regardless of feedback length.
% As such, we used `succinct' feedback when examining newer and, in some cases, more expensive models like GPT-4o.

Results are tabulated as the percent of designs from each test set which were generated successfully, i.e. passing all tests.
\Cref{tab:feedback-successes} gives the success percentages for each combination of LLM, feedback style, number of candidates, and maximum tree search depth for tests with feedback, while~\Cref{tab:zero-shot-successes} gives these results for cases with no feedback.

\begin{table}[!h]
    
\centering
\caption{Percent of passing benchmark designs from VerilogEval-Machine and VerilogEval-Human using AutoChip. Results were gathered with Claude 3 Haiku (Haiku), GPT-3.5-Turbo (GPT-3.5T), GPT-4o-Mini, and GPT-4o. Candidate values of $k=1$ are representative of the original iterative approach to AutoChip.}
\label{tab:feedback-successes}

% \resizebox{0.49\linewidth}{!}{
\begin{tblr}{
colspec = c|c|c|ccc|ccc,
row{1-2} = {font=\bfseries},
column{1} = {font=\bfseries}
}
\toprule
\SetCell[r=2]{c}{Feedback} & \SetCell[r=2]{c}{Candidates} & \SetCell[r=2]{c}{LLM} & \SetCell[c=3]{c}{Eval-Machine (\%)} &&& \SetCell[c=3]{c}{Eval-Human (\%)} && \\ \hline
& &  & d=1 & d=5 & d=10 & d=1 & d=5 & d=10 \\ \hline
\midrule %%%%%%%%%%%%%%%%%%%%%%%%%%%%%%%%%%%%%%%%%%%%%%%%%%%%%%%%%%%%%%%%%%%%%%%%%%%%%%%%%%%%%%%%%%%%%%%
\SetCell[r=4]{c}{Full} & \SetCell[r=2]{c}{$k=1$} & Haiku & 74.1 & 77.6 & 79.7 & 59.6 & 59.6 & 60.2  \\
& & GPT-3.5T & 62.9 & 70.6 & 72.7 & 40.4 & 44.8 & 48.1  \\ \hline
& \SetCell[r=2]{c}{$k=5$} & Haiku & 81.1 & 82.5 & 83.9 & 67.3 & 70.5 & 70.5  \\
& & GPT-3.5T & 79.7 & 86.0 & 86.0 & 52.6 & 59.0 & 64.7  \\ \hline
\hline %%%%%%%%%%%%%%%%%%%%%%%%%%%%%%%%%%%%%%%%%%%%%%%%%%%%%%%%%%%%%%%%%%%%%%%%%%%%%%%%%%%%%%%%%%%%%%%
\SetCell[r=8]{c}{Succinct} & \SetCell[r=4]{c}{$k=1$} & Haiku & 74.1 & 75.5 & 79.7  & 58.3 & 59.6 & 62.8 \\
& & GPT-3.5T  & 62.9 & 71.3 & 72.7  & 42.9 & 49.3 & 52.6  \\
& & GPT-4o-Mini & 68.5 & 69.9 & 69.2 & 55.1 & 60.3 & 60.9 \\
& & GPT-4o  & 77.6 & 81.8 & 84.6  & 64.7 & 72.4 & 75.6  \\
\hline
& \SetCell[r=4]{c}{$k=5$} & Haiku  & 81.8 & 83.2 & 83.9  & 67.9 & 69.2 & 71.8  \\
& & GPT-3.5T  & 81.8 & 86.0 & 86.0 & 58.3 & 58.3 & 66.7  \\
& & GPT-4o-Mini & 73.4 & 74.1 & 76.2 & 64.1 & 65.4 & 71.1 \\
& & GPT-4o & 83.9 & 86.7 & 87.4 & 72.4 & 79.5 & 84.0  \\
\bottomrule

\end{tblr}
% }

\end{table}
\begin{table}[!h]
    \centering
\caption{Zero-shot results for Claude3 Haiku, GPT-3.5 Turbo, and GPT-4o. Results shown for VerilogEval and RTLCoder are reported data for those models, so higher values for k have not been evaluated.}
\label{tab:zero-shot-successes}
%\resizebox{\linewidth}{!}{
\begin{tblr}{
colspec = c|cccccc|cccccc,
row{1-2} = {font=\bfseries},
column{1} = {font=\bfseries}
}
\toprule
\SetCell[r=2]{c}{Model} & \SetCell[c=6]{c}{Eval-Machine (\%)} &&&&&& \SetCell[c=6]{c}{Eval-Human (\%)} &&&&& \\ \hline
& k=1 & k=5 & k=10 & k=25 & k=30 & k=55 & k=1 & k=5 & k=10 & k=25 & k=30 & k=55 \\ \hline
\midrule %%%%%%%%%%%%%%%%%%%%%%%%%%%%%%%%%%%%%%%%%%%%%%%%%%%%%%%%%%%%%%%%%%%%%%%%%%%%%%%%%%%%%%%%%%%%%%%
Haiku & 69.9 & 79.0 & 83.2 & 83.9 & 83.9 & 84.6 & 51.9 & 62.2 & 67.3 & 69.2 & 70.5 & 73.1 \\
GPT-3.5T & 53.1 & 79.7 & 78.3 & 81.8 & 87.4 & 85.3 & 31.4 & 49.4 & 56.4 & 63.5 & 61.5 & 66.0  \\
GPT-4o-Mini & 62.2 & 72.0 & 72.7 & 76.9 & 76.2 & 76.2 & 51.9 & 59.6 & 64.1 & 67.9 & 66.7 & 67.9 \\
GPT-4o & 65.7 & 74.1 & 76.2 & 78.3 & 79.7 & 83.2 & 61.5 & 69.9 & 75.0 & 76.9 & 78.2 & 78.2 \\ \hline \hline
VerilogEval* & 46.2 & 67.3 & 73.7 & - & - & - & 28.8 & 45.9 & 52.3 & - & - & -\\
RTLCoder* & 61.2 & 76.5 & 81.8 & - & - & - & 41.6 & 50.1 & 53.4 & - & - & -  \\
\bottomrule

\end{tblr}
%}
\end{table}

We can further examine these results from two perspectives: model effort, estimated by the average input and output tokens needed to generate a successful design, and model complexity, estimated by the \newt{}{U.S. dollar} (\$ USD) costs needed to generate a successful design \newt{}{at the time of experimentation}. 

\begin{figure}[h]
    \centering
    \subfloat[][Success rate for the Eval-Machine benchmark problems given the total number of tokens used.]{
        \includegraphics[width=\linewidth]{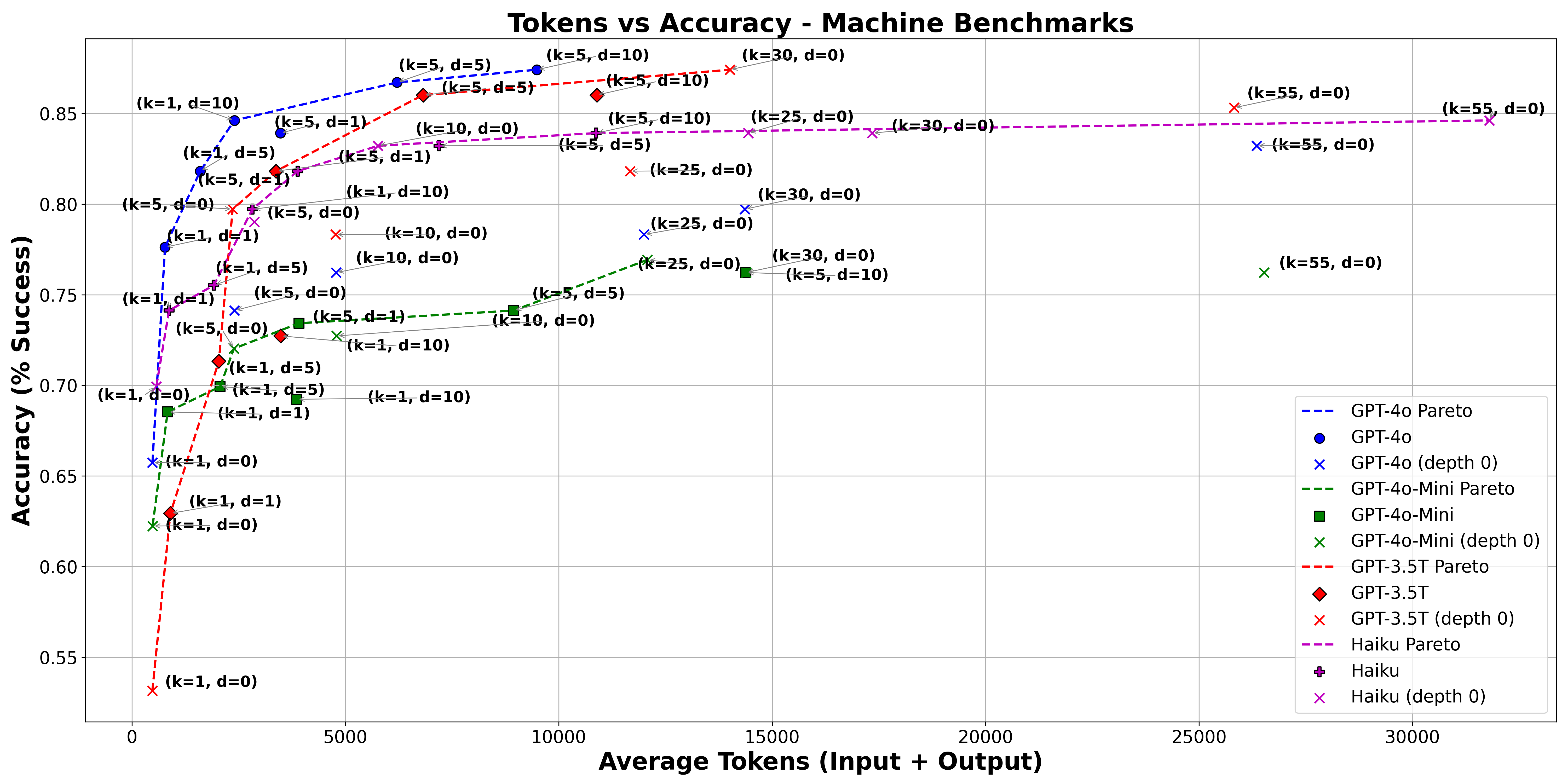}
        \label{fig:tokens-machine-pareto}
    }\\ %
    \subfloat[][Success rate for the Eval-Human benchmark problems given the total number of tokens used.]{
        \includegraphics[width=\linewidth]{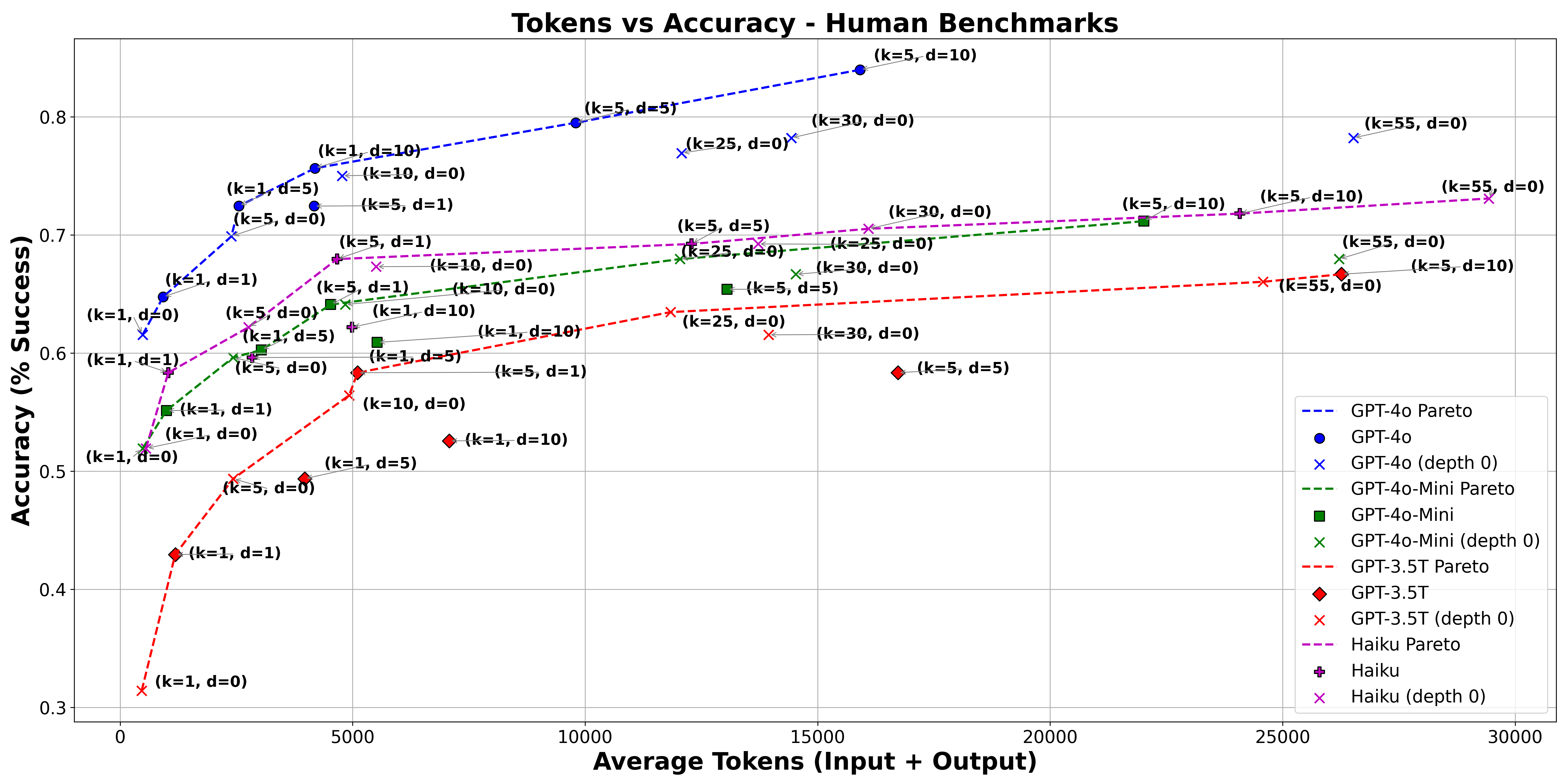}
        \label{fig:tokens-human-pareto}
    }
    \caption{Generated circuit success rates for each evaluated model, shown as the percentage of generated circuits which pass all tests given the average number of tokens needed. Pareto points are plotted with dashed lines. Higher ``accuracy'' at lower ``average tokens'' is better. Data points represented by `x' indicate results from zero-shot testing.}
	\label{fig:tokens-pareto}
\end{figure}

\textbf{Model Effort:}~\Cref{fig:tokens-machine-pareto} and~\Cref{fig:tokens-human-pareto} present an analysis of the proportion of successfully generated benchmark designs based on the average number of tokens (both input and output) needed to generate the best design, on the VerilogEval-Machine and VerilogEval-Human benchmarks respectively.
This metric serves as a proxy for the amount of computational work needed to be done by each model to come up with the best solution.
The Pareto points for each model are connected via dashed lines, identifying the best results from that model given the average number of tokens used.
Generally speaking, if more tokens were needed then the LLM likely had to provide a greater number of responses, indicating that the model had more difficulty in generating a correct design.

These plots show that the three relatively small models tested, GPT-4o-Mini, GPT-3.5-Turbo, and Claude Haiku, all required more tokens for a given set of $(k,d)$ parameters than GPT-4o and, based on their Pareto points, did not consistently benefit from using compilation feedback from Icarus Verilog or the simulations.
GPT-4o, however, seemed to consistently have higher success when leveraging tool-based feedback; almost all Pareto points are from feedback-driven generation and the average number of tokens used was consistently lower with feedback than with the comparable (same maximum number of queries) zero-shot results.

\begin{figure}[h]
    \centering
    \subfloat[][Success rate for Eval-Machine benchmarks given total cost in USD.]{
        \includegraphics[width=\linewidth]{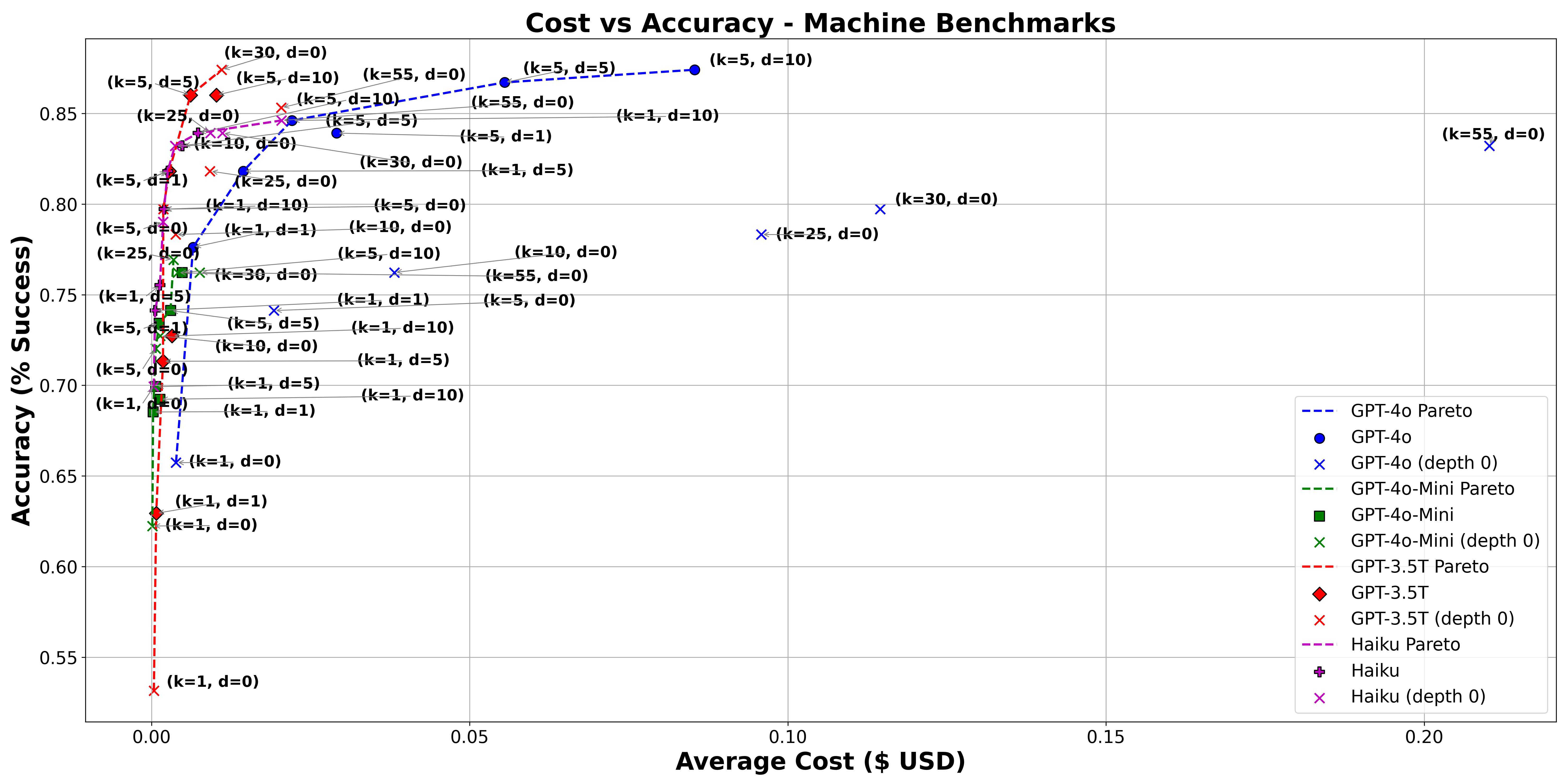}
        \label{fig:cost-machine-pareto}
    }\\ %
    \subfloat[][Success rate for Eval-Human benchmarks given total cost in USD.]{
        \includegraphics[width=\linewidth]{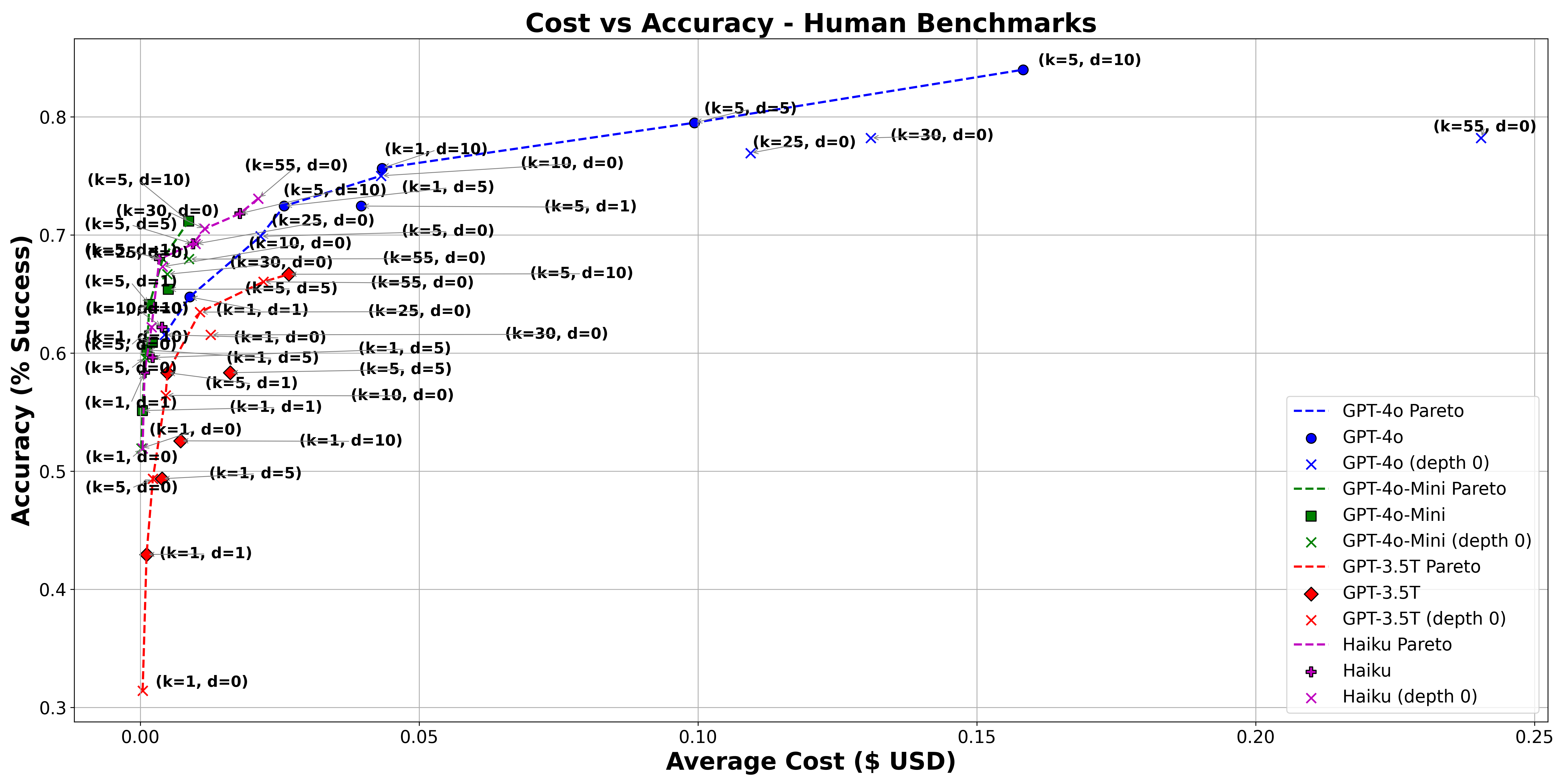}
        \label{fig:cost-human-pareto}
    }
    \caption{Generated circuit success rates for each evaluated model, shown as the percentage of generated circuits which pass all tests given the average USD cost to query the model. Pareto points are plotted with dashed lines. Data points represented by `x' indicate results from zero-shot testing.}
	\label{fig:cost-pareto}
\end{figure}

\textbf{Model Complexity:}~\Cref{fig:cost-machine-pareto} and~\Cref{fig:cost-machine-pareto} show a similar analysis, but rather than giving success as a function of the tokens used, they report success as a function of the \$ USD cost at the time of publication.
We see that, often, more complex models are both more successful and use fewer tokens to reach that level of success; however, these larger models are more computationally complex, so while they use fewer tokens across complete tests, they do far more with those tokens.
This difference in complexity is \newt{reflected in, though we don't know how for certain, the cost to use those models from OpenAI and Anthropic.}{likely reflected by the varying costs for using different models, as determined by each model vendor, though how exactly to quantify this remains an open challenge}. 
While the relationship between model complexity and cost charged is unknown, it stands to reason that the more complex and capable models would cost more.

The costs of the models evaluated in this paper are given in~\Cref{tab:evaluated_llms}, and are used along with the token usage data to determine the average functional cost of generating a design for each benchmark.
While GPT-4o had the highest rate of success and required the fewest average number of tokens for that success, the cost of these tokens far outstrips any of the smaller models tested.
The cost of generation with tool feedback, though, was still significantly lower than with equivalent maximum potential candidates evaluated as zero-shot (\textbf{RQ3}).

% To compare the results due to feedback to the zero-shot values, we examine the success based on the average number of tokens, both input and output, necessary to achieve the highest success rate for each benchmark.
% {\color{red}More tokens being required to solve a problem implies more computational cost...}
% These actual counts of LLM queries provide insight into the real effort required to generate correct HDL, as opposed to the potential maximum effort required.

% For each model tested, the average number of tokens per test parameter setup was calculated and plotted against the proportion of successful designs from the relevant benchmarking set.

\textbf{Impact of Feedback:} Both~\Cref{fig:tokens-pareto} and~\Cref{fig:cost-pareto} show that feedback \textit{can} improve the quality of code given model effort and complexity, however it depends largely on the model being used---it is not a given (\textbf{RQ1}).
This is particularly the case with GPT-4o, the most complex model we tested, where the presence of feedback consistently improved correctness rates.
This may indicate that more capable models are better able to extrapolate the causes of design and implementation bugs given errors, and other suitably large models may also be able to benefit from tool-based feedback.

\textbf{Impact of Tree Search Parameters $(k,d)$:} For our analysis, we employ a similar ``Pass@k'' metric to that used in prior works on this topic, like VerilogEval~\cite{liu_verilogeval_2023} and RTLCoder~\cite{liu_rtlcoder_2024}, which refers to the number of candidates generated for a prompt and considering a circuit a success if at least one candidate proved to work.
Our $k$ candidates to a single prompt functions similarly with a tree search depth $d$ of 0; however, analyzing by the potential maximum number of LLM responses, as the zero-shot Pass@k metric does, we instead use $k * (d+1)$.
With this comparison, we find that increasing both $k$ and $d$ separately improves the resulting circuits, though $k$ seems to have a larger impact (\textbf{RQ2}).
% This is, however, not entirely consistent across the models tested.

The rate of improvement with number of candidates and depth, both separately and when considered together, seems to slow down significantly as the maximum depth increases, likely due to the fact that most simple problems are solved with fewer iterations and only a small handful of the benchmarking problems that can be solved by this method will be done with a larger number of iterations.

% Increasing the maximum depth of the tree search improves the success rate of the benchmarks.
% However, the rate of improvement is non-linear and slows down as the maximum depth increases, indicating that improvement might approach a maximum value where additional effort does not improve the success (\textbf{Ans RQ3}).
%  Other works on this topic, such as RTLCoder~\cite{liu_rtlcoder_2024} and VerilogEval~\cite{liu_verilogeval_2023} use a ``Pass@k'' metric to evaluate the language models.
% This refers to generating $k$ responses per LLM prompt, and considering the result successful if at least one is deemed correct.
% We refer to multiple responses to a single prompt as ``candidates,'' so examining AutoChip's results for $k$ candidates at a search depth of 0 gives an equivalent metric for comparison.
% \Cref{tab:zero-shot-successes} shows our proportion of successful answers for tests with depth 0, as well as the best reported results from both RTLCoder and VerilogEval where provided.
% %For this comparison we examine the results from RTLCoder's best performing model, RTLCoder-DeepSeek.
% All of the models tested show improved accuracy given more candidate responses, but the average number of LLM calls necessary to reach a similar accuracy to a design with feedback is greater.

\begin{figure}[!h]
    \centering
    \subfloat[][Eval-Machine benchmark problem success, separated by major category used by HDLBits.]{
        \includegraphics[width=\linewidth]{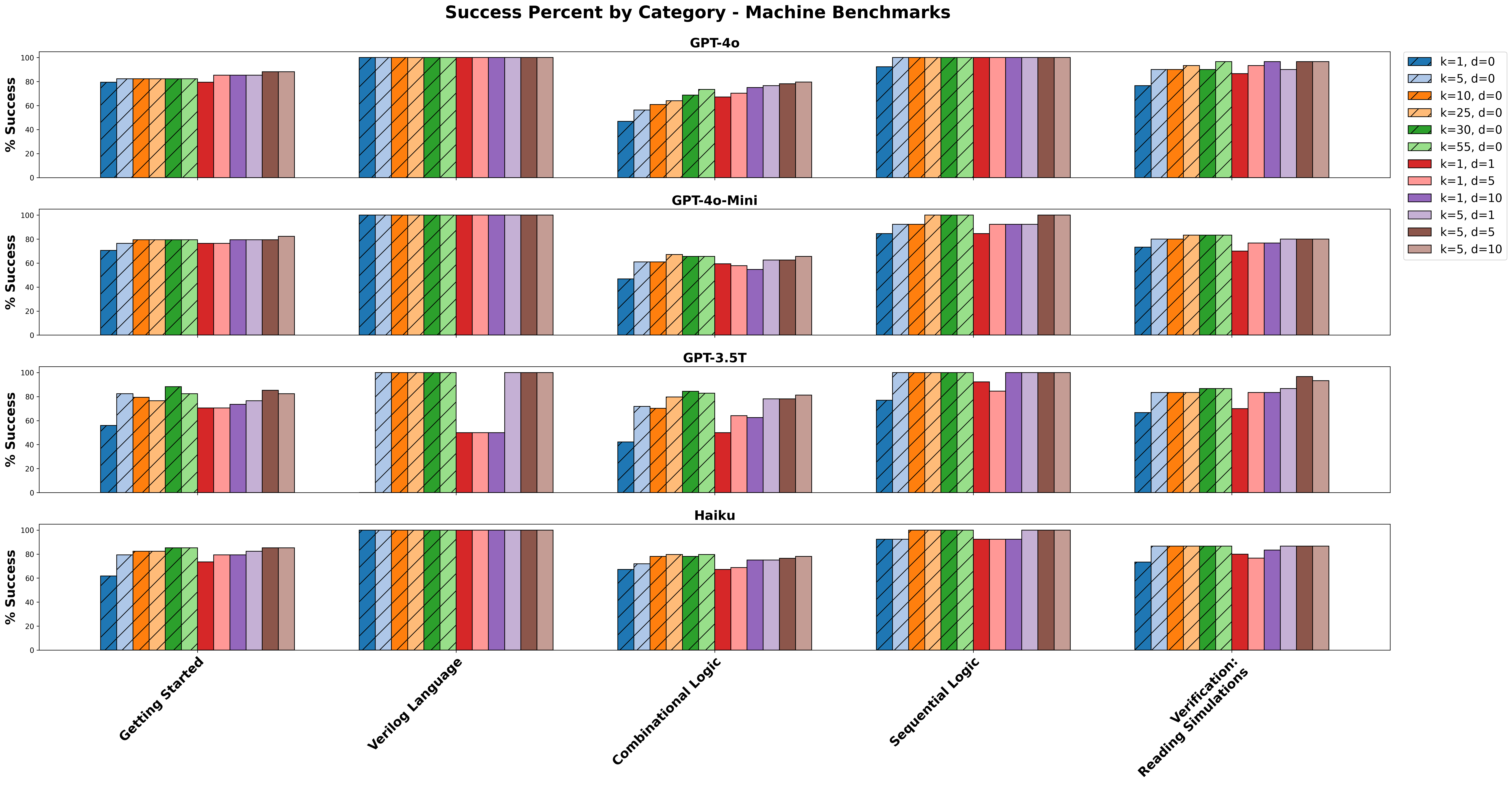}
        \label{fig:categories-machine}
    }\\
    \subfloat[][Eval-Human benchmark problem success, separated by major category used by HDLBits.]{
        \includegraphics[width=\linewidth]{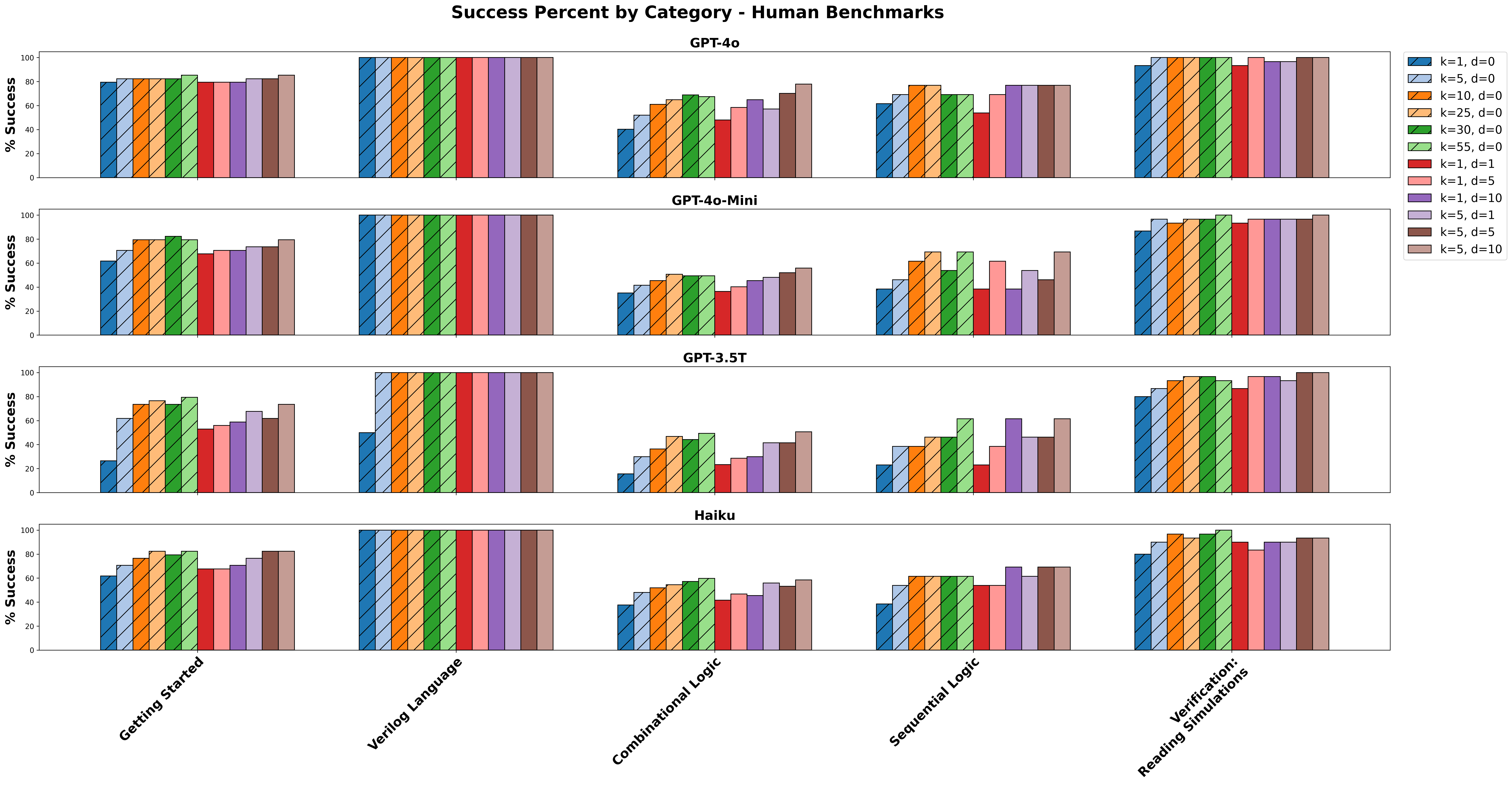}
        \label{fig:categories-human}
    }
    \caption{Model success, separated by category.  Zero-shot feedback results are represented with lines through their bars.}
    \label{fig:categories}
\end{figure}

\textbf{Benchmark Categorization:} HDLBits, and subsequently its problems used by VerilogEval, sorts its problems into categories and subcategories, ranging from ``Getting Started'' and ``Verilog Language,'' which consist of simple problems focusing on specific features of Verilog and their syntax, up to ``Sequential Logic'' and ``Verification: Reading Simulations,'' which ask for generally more complex or abstractly worded problems dealing with state machines and sequential functions, and reading waveforms, respectively.
This problem categorization enables us to analyze the performance of each model we tested on different kinds of Verilog problem and identify patterns in what models handle what problems well, and how well the tool-based feedback is able to improve performance per-category.
\Cref{fig:categories-machine} and~\Cref{fig:categories-human} show the success rates of each examined model, given the major HDLBits categories and feedback configurations discussed above.

Across all models we found that ``simple'' problems, such as basic Verilog functionality questions, are able to be solved more often, regardless of feedback depth or number of candidates.
There is a general trend where more iterations of feedback or more candidates improves results, but this is to be expected given our previous results where an increased number of queries often results in higher success, regardless of the parameter configuration for those queries (zero-shot or tree search).
We don't identify any particular categories of problem which seem to benefit from using tool feedback, even in the case of a model that generally seems to benefit like GPT-4o.

Breaking down the benchmarking problems by category also allows us to better examine the differences between the results from the VerilogEval-Machine benchmarks and the VerilogEval-Human benchmarks.
Following the combined results from above, the rate of success for the machine benchmarks appears to be higher across the board; however, the Machine benchmark ``Sequential Logic'' problems had significantly more success than their combinational counterparts, whereas the Human benchmark set's ``Sequential Logic'' problems had roughly similar results to their accompanying combinational problems.
This indicates that there may be more successful prompting strategies depending on the specific problem trying to be solved.

We can further decompose the problem categories into their, once again HDLBits-defined, subcategories.
These better separate the relative difficulties and styles of problem.
For instance, both individual latch and flip-flop problems and cellular automata problems fall under ``Sequential Logic'', but the latter are more comprehensive problems with abstract wording.
As such, the subcategories give finer-grained insight into the types of problems different models are suited to solve.
% These categories are shown in~\Cref{tab:categories}; however, they have been slightly modified.
% Some subcategory names were changed to be more descriptive of the problems they contain, and a supercategory, CS450, had its benchmarks placed in other categories which were more descriptive of the types of design being requested.
%This categorization provides insight into the types of problem which LLMs are best suited to solving and trends in how to generate the most successful results.

\begin{sidewaysfigure}
    \centering
    \subfloat[][Eval-Machine benchmark problem success, separated by subcategory used by HDLBits.]{
        \includegraphics[width=\linewidth]{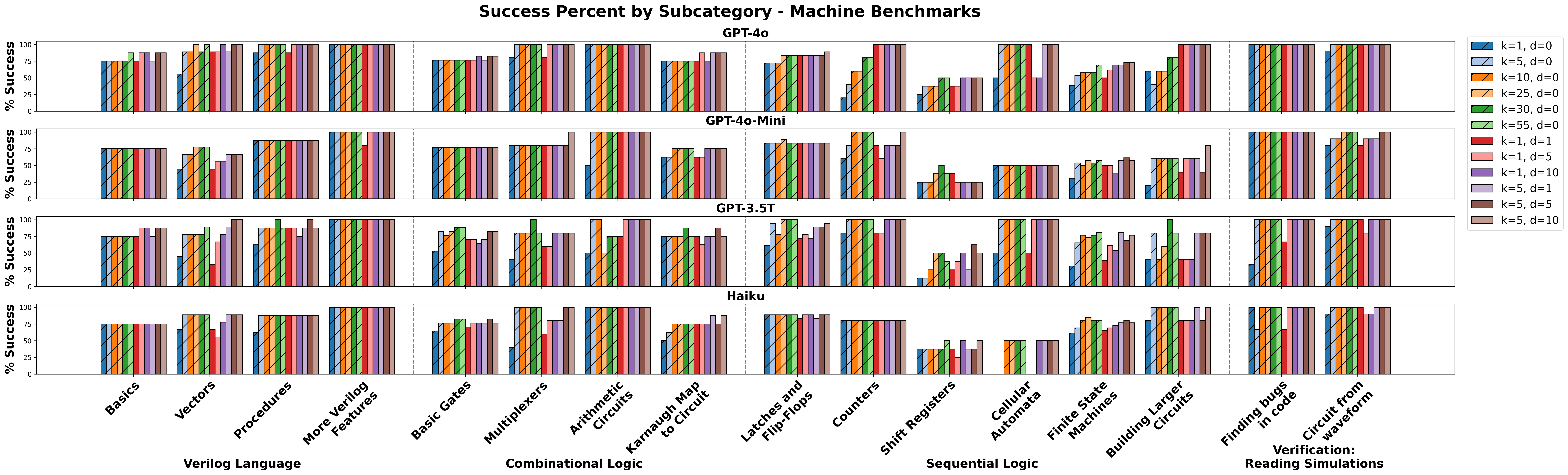}
        \label{fig:subcategories-machine}
    }\\
    \subfloat[][Eval-Human benchmark problem success, separated by subcategory used by HDLBits.]{
        \includegraphics[width=\linewidth]{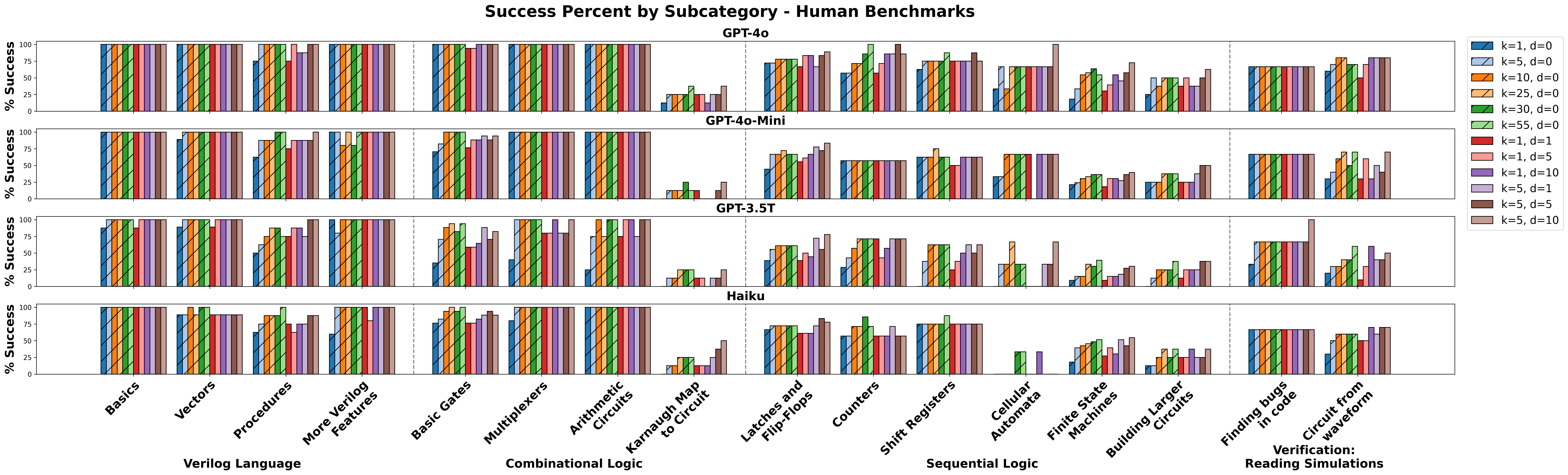}
        \label{fig:subcategories-human}
    }
    \caption{Problem success rates by problem subcategory for both the Eval-Machine and Eval-Human problem sets. The major category to which the subcategories belong is given below each group.}
    \label{fig:subcategories}
    \vspace{-450pt}
\end{sidewaysfigure}

\Cref{fig:subcategories-machine} and~\Cref{fig:subcategories-human} show the rate of success on the benchmark set for each parameter set and each model, broken down by subcategory, for both the Eval-Machine and Eval-Human benchmark sets.
Some problems, such as the ``Getting Started'' category, are omitted from this graph as they contained too few and too simple problems to provide useful insight.

We observe that the simplest problem types are often solved with few candidate responses, not necessarily requiring feedback from the tools, but more complex and abstract problems benefit more from a tool feedback-driven approach.

The problems for the Human dataset which the models seemed to struggle with the most were interpreting Karnaugh maps (provided as text), state machine design, and problems that are presented as being more ``abstract'' such as the cellular automata problems.
For the Machine dataset, which uses far more straightforward prompts, the LLMs had consistently lower success with simple designs, such as the ``Verilog Language'' category, and ``Basic Gates'' and ``Multiplexers'' from ``Combinational Logic,'' but showed much greater success with  complex designs such as  state machines or finding bugs (\textbf{RQ5}).
% While the overall success rate is consistently high for the Machine problem set, this  does not hold true across all problem categories, potentially indicating that the prompting strategy should be tailored to the circuit being generated.

%\begin{landscape}
%\end{landscape}

\subsection{Mixed-Model Results}
Given the relative success, but matching relative expense, of GPT-4o in solving the VerilogEval benchmarking problems, we sought to evaluate if combining a small model with a larger model, such as GPT-4o, could achieve improved results with only minimal impact on cost.
\Cref{tab:feedback-successes-ensemble} shows the results for using Claude Haiku, GPT-3.5-Turbo, and GPT-4o-Mini \newt{}{for the complete tree search} with GPT-4o as \newt{}{only} the final iteration \newt{}{in cases where the maximum depth was reached}.
\begin{table}[h]
    
\centering
\caption{Percent of passing benchmark designs from VerilogEval-Machine and VerilogEval-Human using AutoChip with mixed-models. Results were gathered with Claude 3 Haiku (Haiku), GPT-3.5-Turbo (GPT-3.5T), and GPT-4o-Mini, each followed by a single pass of GPT-4o as the final iteration.}
\label{tab:feedback-successes-ensemble}

%\resizebox{\linewidth}{!}{
\begin{tblr}{
colspec = c|c|ccc|ccc,
row{1-2} = {font=\bfseries},
column{1} = {font=\bfseries}
}
\toprule
\SetCell[r=2]{c}{Candidates} & \SetCell[r=2]{c}{LLM} & \SetCell[c=3]{c}{Eval-Machine (\%)} &&& \SetCell[c=3]{c}{Eval-Human (\%)} && \\ \hline
&  & d=1 & d=5 & d=10 & d=1 & d=5 & d=10 \\ \hline
\midrule %%%%%%%%%%%%%%%%%%%%%%%%%%%%%%%%%%%%%%%%%%%%%%%%%%%%%%%%%%%%%%%%%%%%%%%%%%%%%%%%%%%%%%%%%%%%%%%
\SetCell[r=3]{c}{$k=1$} & Haiku & 75.5 & 79.0 & 81.1 & 64.7 & 67.3 & 67.9 \\
& GPT-3.5T  & 77.6 & 80.4 & 78.3 & 62.2 & 63.5 & 61.5 \\
& GPT-4o-Mini & 69.9 & 74.8 & 76.9 & 63.5 & 64.1 & 66.0 \\
\hline
\SetCell[r=3]{c}{$k=5$} & Haiku  & 86.0 & 86.7 & 85.3 & 74.4 & 75 & 74.4  \\
& GPT-3.5T  & 86.7 & 86.7 & 86.0 & 71.2 & 73.7 & 71.8 \\
& GPT-4o-Mini & 82.5 & 81.1 & 81.8 & 69.9 & 74.3 & 75.0 \\
\bottomrule

\end{tblr}
%}

\end{table}

These results show significant promise when adding a more complex model to the end of a series of queries from less capable models, with the rate of success shown to approach or even exceed, in some cases, the benchmark circuits generated using only GPT-4o.

\begin{figure}[h]
    \centering
    \subfloat[][Mixed-model success rates for the Eval-Machine benchmark problems given the total number of tokens used.]{
        \includegraphics[width=\linewidth]{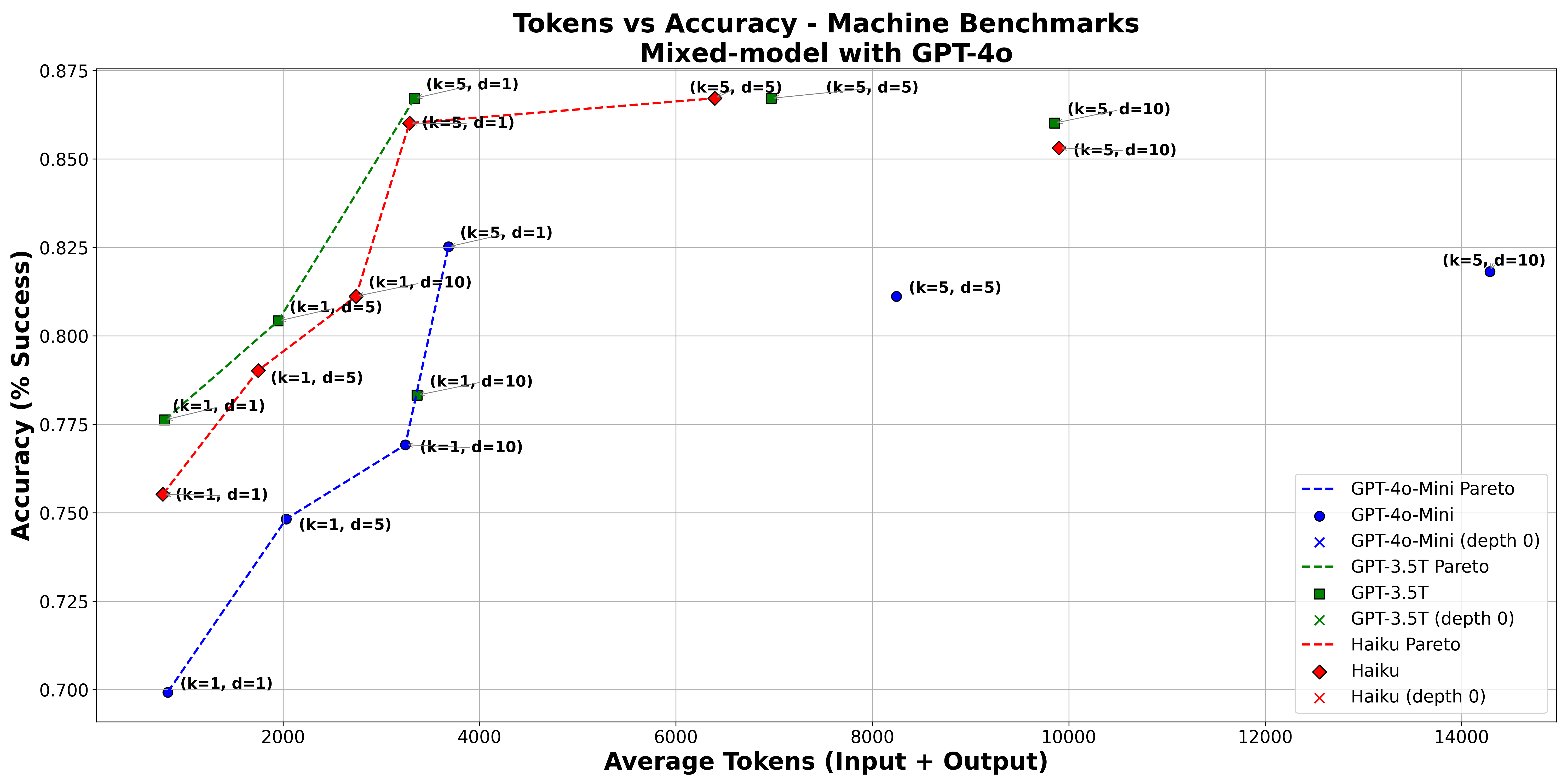}
        \label{fig:tokens-ensemble-machine}
    }\\ %
    \subfloat[][Mixed-model success rates for the Eval-Human benchmark problems given the total number of tokens used.]{
        \includegraphics[width=\linewidth]{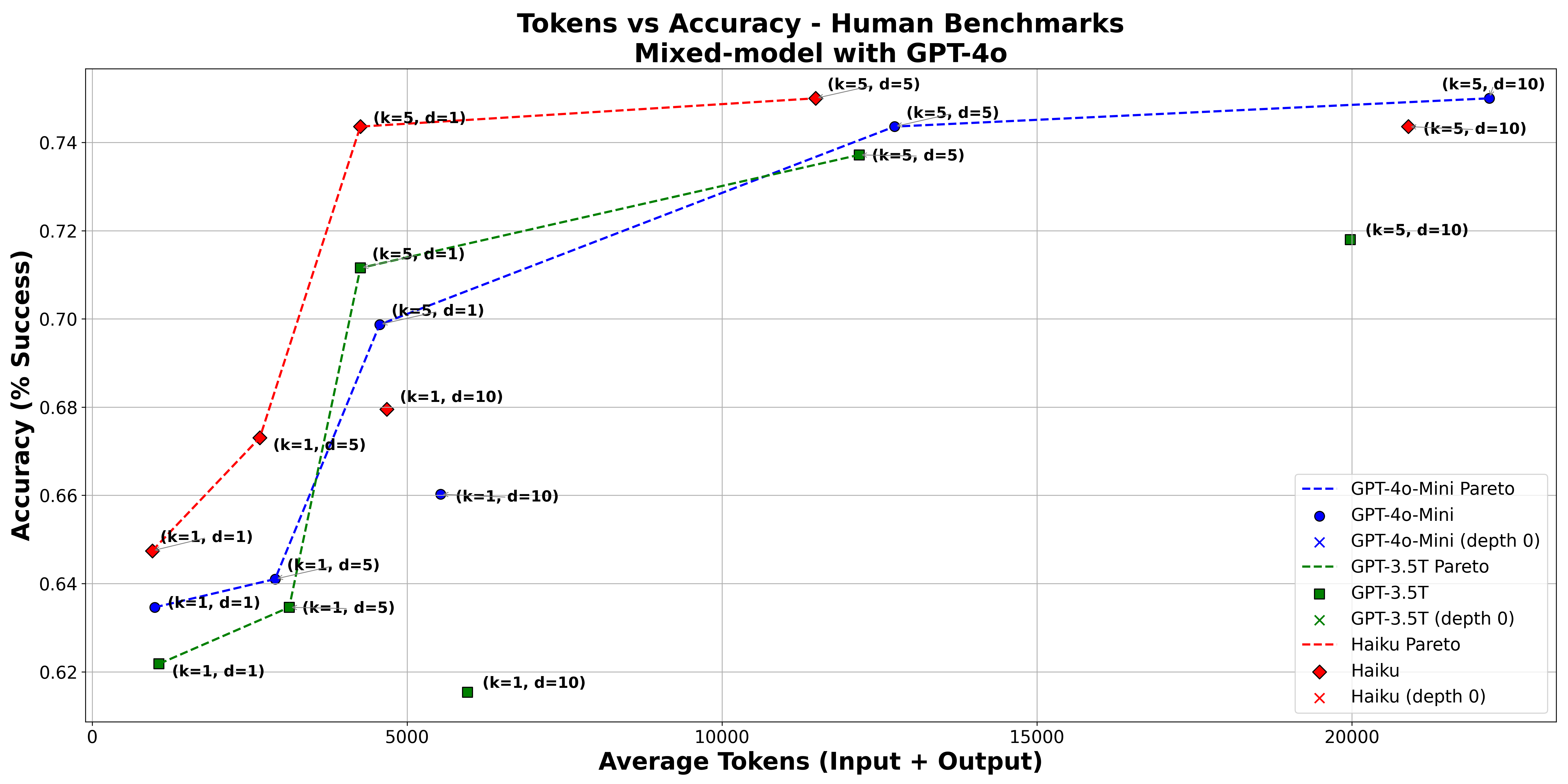}
        \label{fig:tokens-ensemble-human}
    }
    \caption{Generated circuit success rates for each evaluated model, shown as the percentage of generated circuits which pass all tests given the average number of tokens needed. Each model's final iteration was done with GPT-4o. Pareto points are plotted with dashed lines.}
\end{figure}

Evaluating these results based on model effort, as with the single-model results, is done by analyzing the average number of tokens needed to completely generate a benchmark design.
\Cref{fig:tokens-ensemble-machine} and~\Cref{fig:tokens-ensemble-human} show these plots.

We see that with the VerilogEval-Machine benchmark sets we are using a similar number of tokens to the small models' expenditures alone, while achieving notably higher levels of success.
We also observe that the Pareto points in most cases seem to stop before the largest combinations of candidates and tree search depth, indicating that by mixing these models smaller search parameters can be used to still achieve a relatively high rate of success.
With the more realistic VerilogEval-Human benchmarks, we are not able to reach the same level of success as GPT-4o on its own, but we do see notably higher success rates compared to the smaller models for similar numbers of tokens.

\begin{figure}[h]
    \centering
    \subfloat[][Mixed-model success rate for Eval-Machine benchmarks given total cost in USD.]{
        \includegraphics[width=\linewidth]{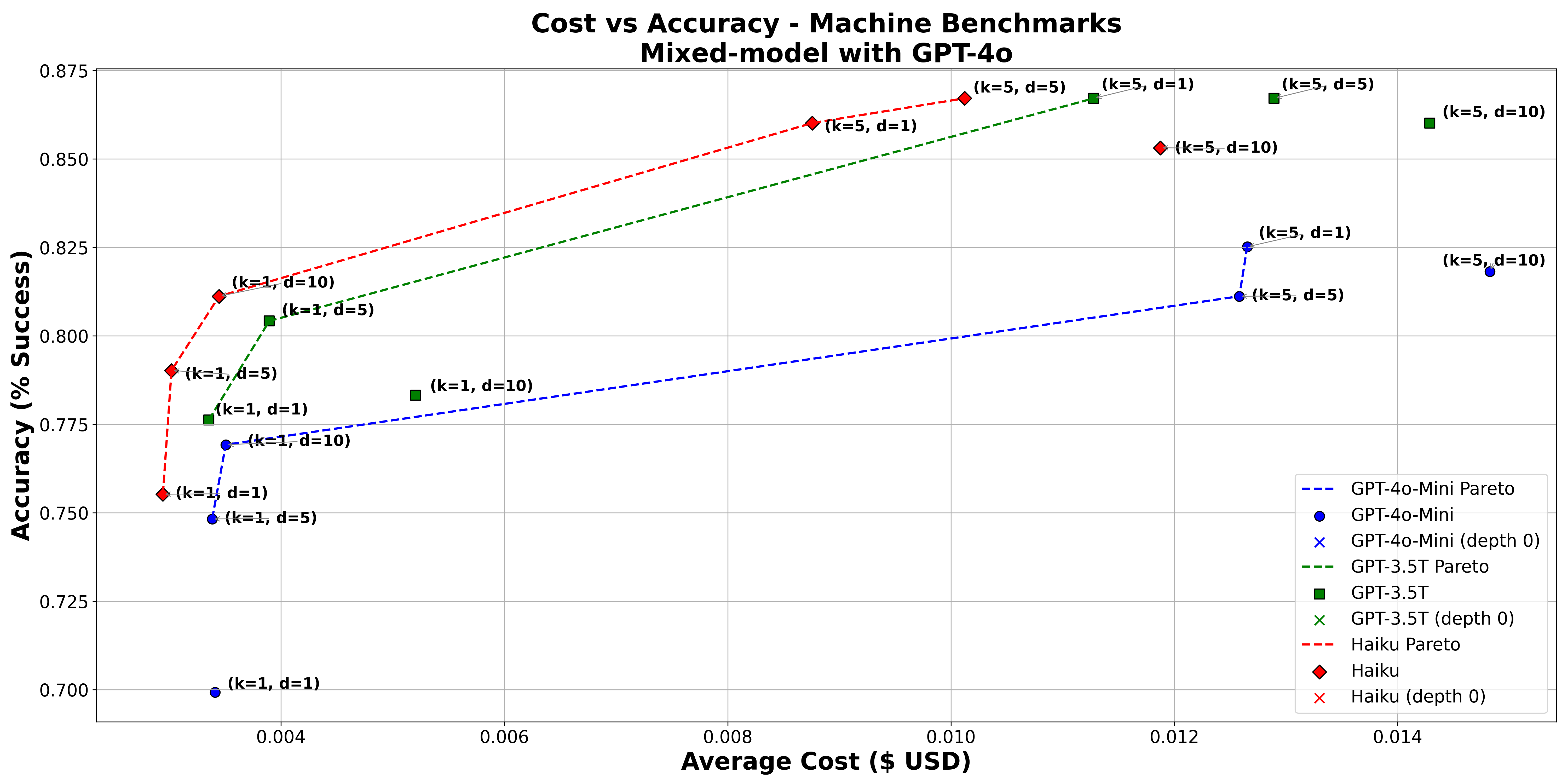}
        \label{fig:cost-ensemble-machine}
    }\\ %
    \subfloat[][Mixed-model success rate for Eval-Human benchmarks given total cost in USD.]{
        \includegraphics[width=\linewidth]{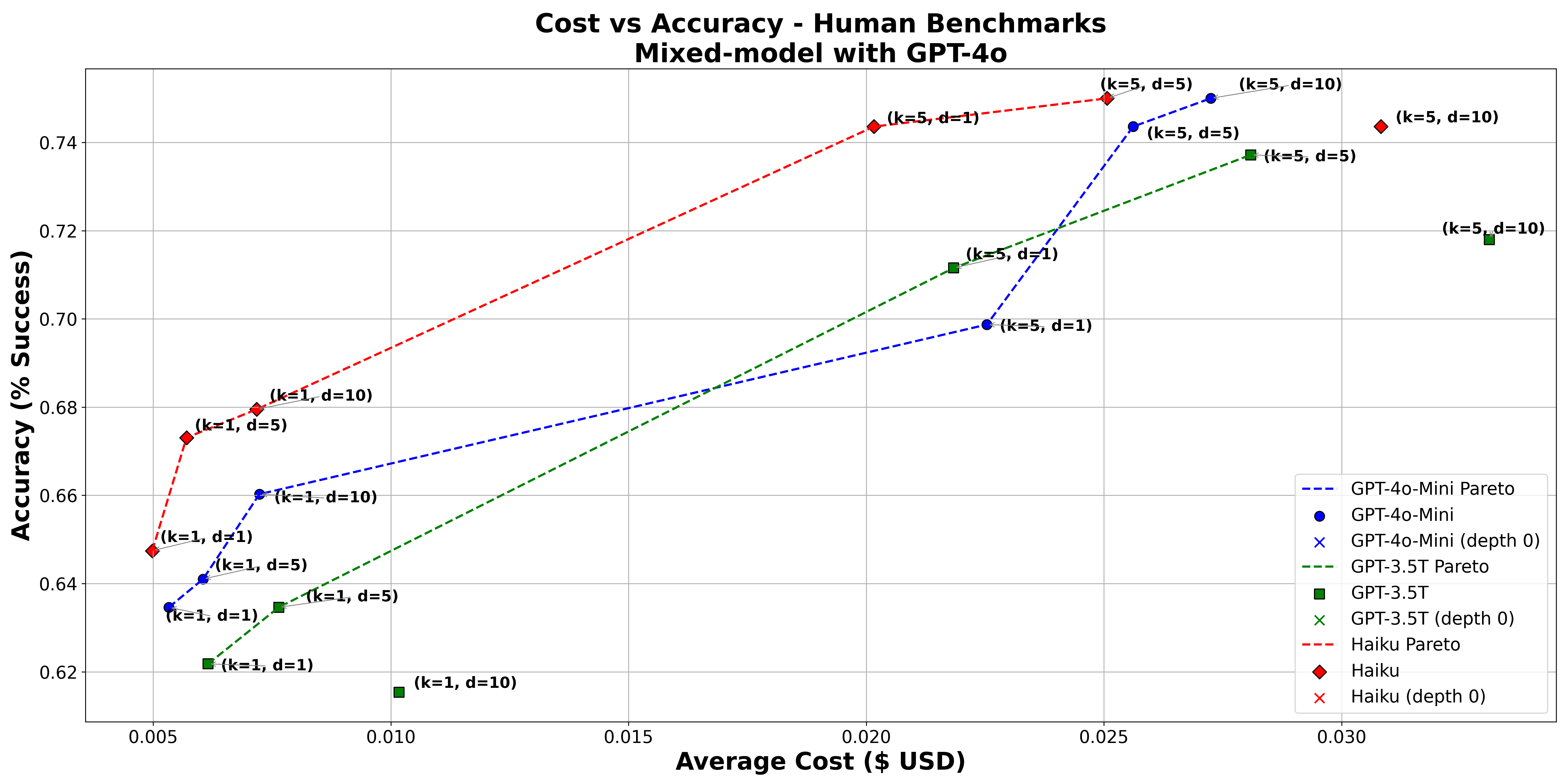}
        \label{fig:cost-ensemble-human}
    }
    \caption{Generated circuit success rates for each evaluated model, shown as the percentage of generated circuits which pass all tests given the average USD cost to query the model. Each model's final iteration was done with GPT-4o. Pareto points are plotted with dashed lines.}
\end{figure}

We can also evaluate success based on relative model complexity, once again proxied by the dollar costs necessary to access and run each model.
These plots are shown in~\Cref{fig:cost-ensemble-machine} and~\Cref{fig:cost-ensemble-human}.
Here we observe a similar phenomenon to that of the model effort analysis, but to a greater magnitude.
For VerilogEval-Machine, the average cost to complete the benchmark generation is on average similar, if slightly higher, than the costs for the smaller models on their own, but are orders of magnitude lower than when using GPT-4o alone while achieving similar levels of success.

While VerilogEval-Human does not reach the level of success of GPT-4o alone, we are able to achieve more success than the smaller models could get alone, for a very similar cost.
For example, the highest success rate achieved with mixed-models for the VerilogEval-Human benchmarks was 75\% at a cost of \$0.025 and to achieve a similar nearly 75\% success with only a single model would require GPT-4o and cost \$0.043---an increase of 72\% cost.
This indicates that by mixing models it may be possible to leverage less computational resources to generate Verilog of comparable quality to a computationally heavy model on its own, though prompting method appears to have a significant effect (\textbf{RQ6}).

\begin{figure}[h]
    \centering
    \subfloat[][Success rate of Eval-Machine problems, separated by category.]{
        \includegraphics[width=\linewidth]{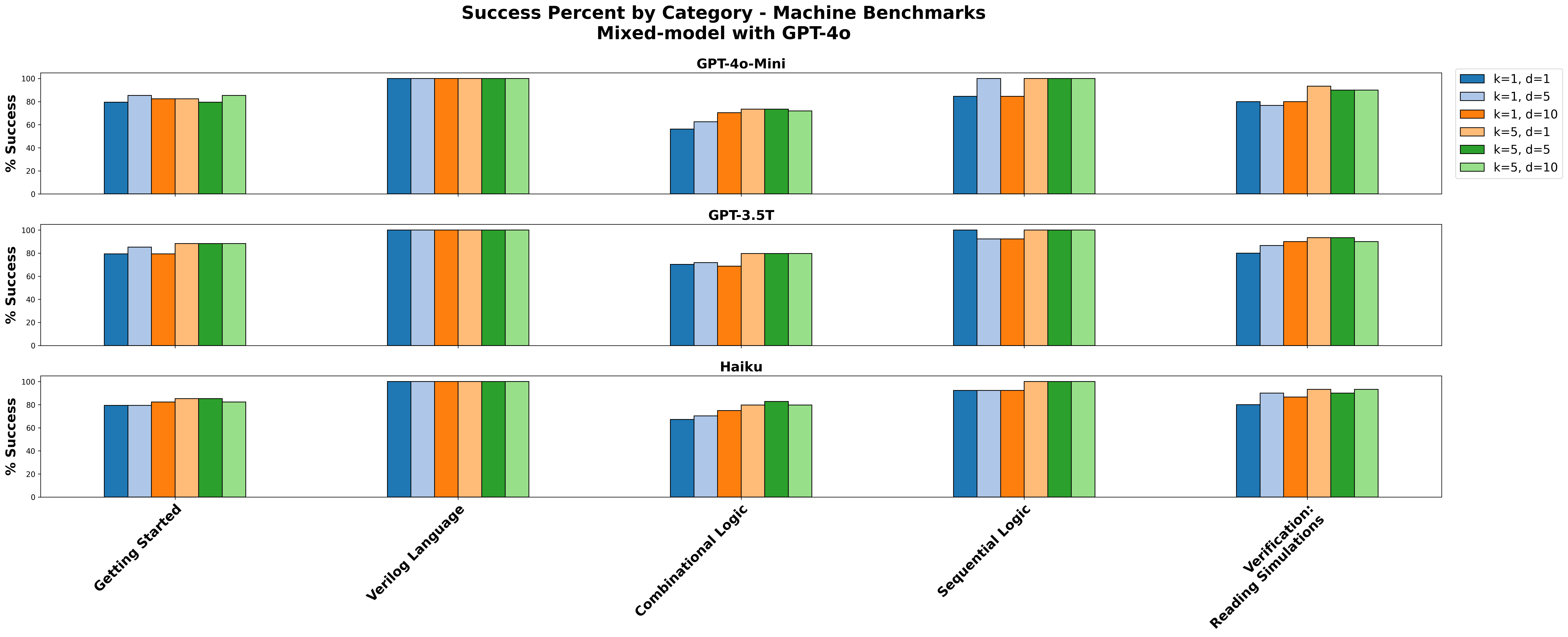}
        \label{fig:categories-ensemble-machine}
    }\\
    \subfloat[][Success rate of Eval-Human problems, separated by category.]{
        \includegraphics[width=\linewidth]{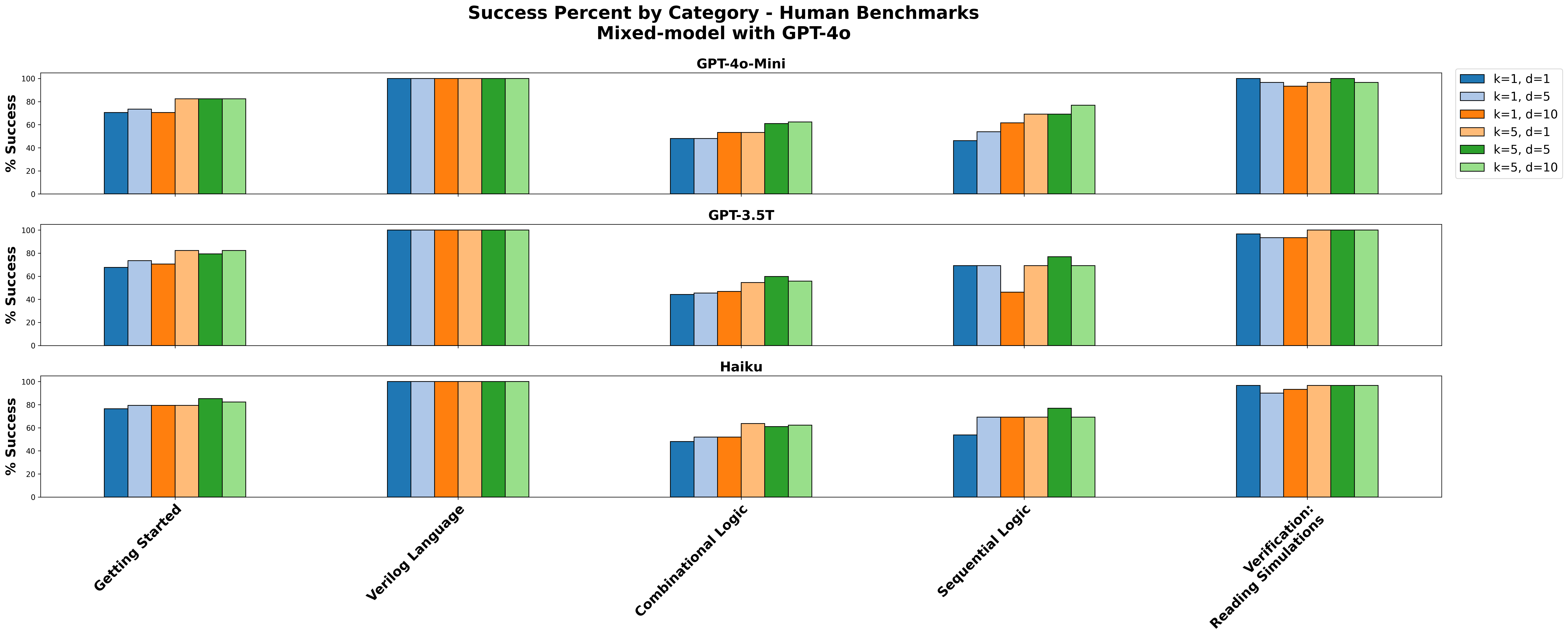}
        \label{fig:categories-ensemble-human}
    }
    \caption{Model successes broken down by category,using GPT-4o for the final iteration of feedback.}
    \label{fig:categories-ensemble}
\end{figure}

\begin{sidewaysfigure}[]
    \centering
    \subfloat[][Success rate of Eval-Machine problems, separated by subcategory.]{
        \includegraphics[width=\linewidth]{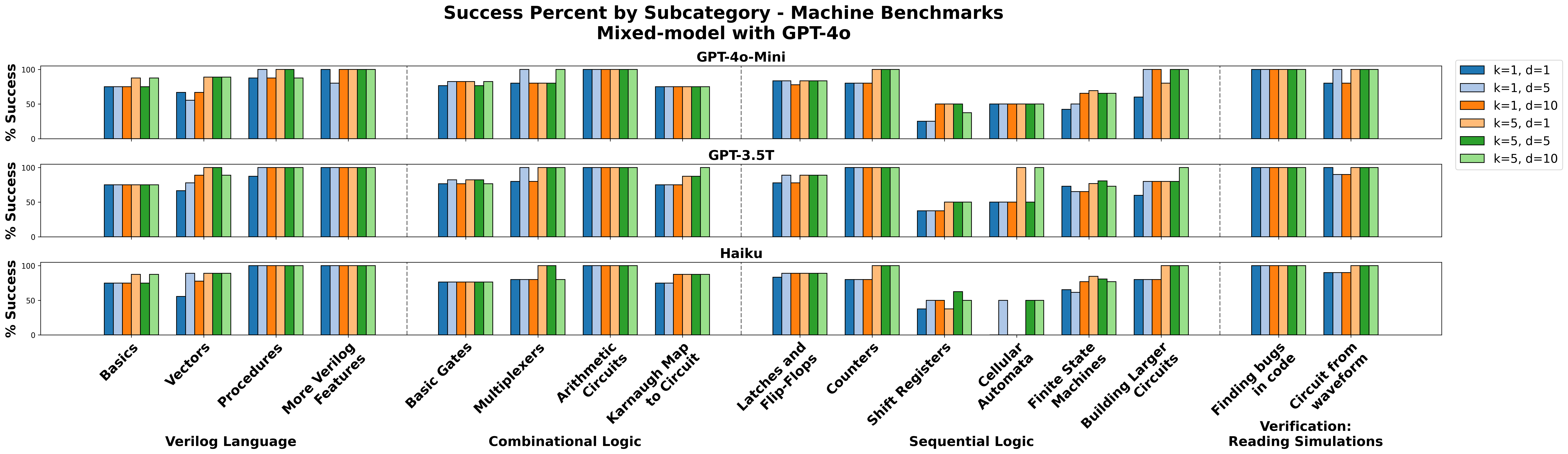}
        \label{fig:subcategories-ensemble-machine}
    }\\ %
    \subfloat[][Success rate of Eval-Human problems, separated by subcategory.]{
        \includegraphics[width=\linewidth]{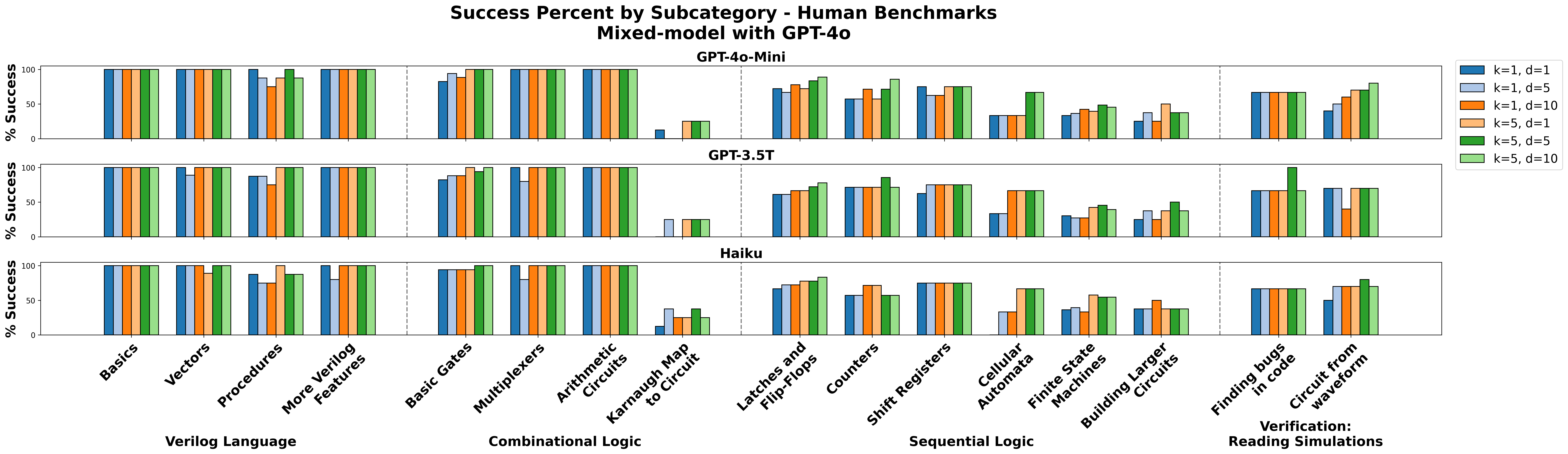}
        \label{fig:subcategories-ensemble-human}
    }
    \caption{Model successes broken down by subcategory, using GPT-4o for the final iteration of feedback.}
    \label{fig:subcategories-ensemble}
    \vspace{-450pt}
\end{sidewaysfigure}

\textbf{Benchmark Categorization:} The results of using multiple models can also be analyzed by category and subcategory.
\Cref{fig:categories-ensemble} and~\Cref{fig:subcategories-ensemble} show the mixed-model results when using the main models shown for most generation, followed by a final iteration with GPT-4o.

Ultimately, we see largely the same pattern of success rates based on category that we observe in the single-model results (\Cref{fig:categories}), but, as noted in our above analysis, those success rates are far closer to GPT-4o's single model results than the smaller models' results.
\Cref{fig:subcategories-ensemble-machine} and~\Cref{fig:subcategories-ensemble-human} once again further break down these categories into their subcategories for a more fine-grained analysis.

%\begin{landscape}
%\end{landscape}
Using mixed-models once again shows that the per-subcategory successes for GPT-4o-Mini, GPT-3.5-Tubro, and Claude 3 Haiku give largely the same pattern of proficiencies, deficiencies, and success rates observed when using only GPT-4o.
This further supports the idea that mixing models can improve the results from the less computationally expensive model while still keeping the expense low.

\subsection{Discussion}
We asked six initial research questions to guide our evaluation of LLM generated Verilog.

% RQ1: Does feedback from hardware verification tools improve LLM-generated HDL over zero-shot results?
\textbf{RQ1:} We found that feedback from tools and testbench simulation \textit{can} improve the success rate of generated Verilog modules over zero-shot results, but it was heavily dependant on the model being used.
GPT-4o consistently benefited from the automated feedback from tools, but the smaller models did not seem to use feedback to repair bugs as well.
This could be an indication of the complexity of relating error messages and simulation times to the design being created, a similar problem to one found when using different levels of feedback from humans~\cite{blocklove_evaluating_2024}.
In the previous work, a human engineer was able to guide the LLM by explaining, even basically, the correlation between the error and the aspect of the Verilog which caused it.
To improve upon this in an automated system with no human feedback, the LLMs being used would likely need to be previously instructed on these specific error messages.

% RQ2: Does the number of iterations and candidate responses impact quality and number of correct implementations?
\textbf{RQ2:} As the number of candidate responses and the depth of the tree increased we saw a trend higher rates of success with all examined models.
The tree search depth, while impactful, seemed to have a smaller effect on the rate of success than the number of candidates did, as evidenced by the rate of success for zero-shot results with many candidates.
Increasing the depth, however, resulted in generally fewer tokens needed for the increase as compared to increasing the number of candidates.
\newt{}{Future work would seek to better explore the reasons for this and what hyperparamters might impact this outcome, as identifying an ideal number of candidates and tree-search depth would potentially help optimize the efficiency of generating function designs even more.
Further, an examination of multiple tree-search methodologies, such as MCTS, could shed additional light on the most effective ways to use these models, though such an experiment was considered out of scope for this work.}

% RQ3: What is the impact of tool feedback-driven code generation on cost?
\textbf{RQ3:} Leveraging EDA tool feedback with LLMs to improve the generated HDL resulted in consistently lower computational and monetary cost for a given number of queries to the models we evaluated.
We find that the zero-shot results require more tokens and, as such, have more associated expense per model than the results which leveraged feedback from tools.
In the case of GPT-4o, where feedback noticeably improved the results, the lessened cost of using tool feedback given the rate of correct designs was substantial.

% RQ4: Does the amount of context given with feedback have an impact on the rate of successful designs?
\textbf{RQ4:} By using AutoChip with both ``succinct'' and ``full-context'' feedback, we found that there was no appreciable difference in the success rate of the two models we examined, GPT-3.5-Turbo and Calude 3 Haiku.
As such, we used only ``succint'' feedback for testing the other models, as it reaches the same levels of success while using far fewer tokens over the course of larger tree searches.

% RQ5: Are there particular types of hardware design problem which LLMs are more well-equipped to solve than others?
\textbf{RQ5:} The LLMs we examined all behaved consistently with regards to their successes based on the class of problem.
All models seemed to have significant and consistent success with basic features of Verilog, primarily focusing on syntax and simple logical functions.
The models' ability to generate correct designs seemed to generally decline with more complex questions dealing with implementing sequential logic and interpretation of abstract information like Karnaugh maps (K-maps), finite state machine diagrams, and waveform analysis.
The VerilogEval-Machine benchmarks showed noticeably more success with generating circuits based on interpreting design specifications, likely due to the less abstract prompts generated by using an LLM to generate a prompt based only on a final correct circuit.

% RQ6: Can mixing multiple LLMs with different capabilities during a design “run” improve generation quality at reduced cost?
\textbf{RQ6:} We found that mixing models, specifically by adding a single final iteration of the more capable GPT-4o model, resulted in success rates similar to those with only GPT-4o but while requiring only a fraction of the USD cost, which serves as a proxy metric for the computational complexity.
The average number of overall tokens necessary stayed similar to when only using each of the smaller models, as they were being queried the majority of the time, but the final iteration of GPT-4o seemed to often be able to leverage those partially functional designs and provide the final fixes.
This cost reduction is also very likely due to the rate of success of the smaller models on their own.
They are still able to inexpensively solve many of the simpler problems without ever getting to the final depth of the search to call GPT-4o, so the added expense only applied to the hardest to solve problems.

\section{Conclusion}
Given recent advances in LLM capabilities for both coding and hardware design, it has seemed reasonable to assume that providing a model feedback on its generated code would improve its performance. 
%When you ask a model to correct its own mistakes, there have been observed cases where it was able to do so..
However, generating this feedback, and ``explaining'' to the model how and why it is wrong, has typically been done with a human engineer---something costly, nebulous, and potentially slow if the human engineer needs to be able to find and decipher the error on their own before prompting the LLM.
As such, systematic studies in this area have been lacking.

In this work, we therefore sought to evaluate how a set of modern, state of the art, general knowledge LLMs would respond to feedback only from EDA tools and testbenches.
We sought to discover whether LLMs would be able to solve bugs in their own generated designs without the assistance of a knowledgeable human.
We further asked: If so, how much effort would it take?
How much would it cost?
And, are there techniques \newt{we can use to make this process better}{to improve this process}?

Ultimately, we found that the success of using feedback from tools and testbenches to fix generated Verilog designs depended largely on the model being used.
GPT-4o, the most computationally complex model examined, was able to consistently use tool and testbench generated feedback to generate correct designs from the VerilogEval benchmark sets.
The smaller models tested, GPT-4o-Mini, GPT-3.5-Turbo, and Claude 3 Haiku, inconsistently benefited from the tool provided feedback, but, by adding a final evaluation by GPT-4o, were able to create a similar number of correct circuits at a fraction of the cost of GPT-4o being used alone.
We developed and provide AutoChip as an open-source, extendable framework for evaluating LLMs' abilities to generate Verilog and correct their mistakes using the output from tools.
This can be leveraged to perform similar analysis on additional models and new benchmarks as they become available, further building up an understanding for how best to interact with LLMs to generate quality hardware.
Our presented method provides an important and powerful proof-of-concept for effectively utilizing tool feedback with LLMs for the automatic generation of hardware.

% In this work we comprehensively evaluated current state-of-the-art, commercially available conversational LLMs for iterative hardware development with a workflow similar to that which may be undertaken by human engineers.
% We found that iterative feedback via a greedy tree search algorithm, AutoChip, improved the rate of successfully designed circuits given a comparatively small number of required LLM queries.
% AutoChip showed up to 87.4\% success rates when using GPT-4o, suggesting that this framework provides a pathway towards the automatic design of hardware circuits.
% However, the success of LLM-generated HDL, regardless of feedback, appears to be heavily based on the type of problem being presented and the manner used to prompt it.

\bibliographystyle{ACM-Reference-Format}
\bibliography{benhamram}

\end{document}